\documentstyle[preprint,aps,epsfig,floats]{revtex}
\input epsf.tex
\def\DESepsf(#1 width #2){\epsfxsize=#2 \epsfbox{#1}}
\begin{document}
\preprint{\vbox{\hbox{}}}
\draft
\title{
The CP violating phase $\gamma$ \\from global fit
of rare charmless hadronic B decays
}
\author{$^1$X.-G. He, $^1$Y.-K. Hsiao, $^1$J.-Q. Shi,
$^2$Y.-L. Wu and $^2$Y.-F. Zhou}
\address{
$^1$Department of Physics, National Taiwan University,
Taipei\\
$^2$ Institute of Theoretical Physics, Academia Sinica, Beijing}
\date{November, 2000. Revised March, 2001}

\maketitle

\begin{abstract}
We study constraints on the CP violating phase $\gamma$ in the
Kobayashi-Maskawa model
using available experimental data.
We first follow the conventional method to up date the
constraint on $\gamma$ by performing
a $\chi^2$ analysis using data from $|\epsilon_K|$, $\Delta m_{B_{d,s}}$
and $|V_{ub}/V_{cb}|$. We also include the recent information on 
$\sin2\beta$ in the analysis. We obtain the best fit for 
$\gamma$ to be $66^\circ$
and
the 95\% C.L. allowed range to be $42^\circ \sim 87^\circ$.
We then develop a method to carry out a $\chi^2$
analysis based on SU(3) symmetry
using data from $B\to \pi \pi$ and $B\to K \pi$. We also discuss
SU(3) breaking effects from model estimate. We find that present data
on $B\to \pi\pi, K \pi$
can also give some constraint on $\gamma$ although weaker than the earlier
method limited by the present experimental errors.
Future improved data will provide more stringent constraint. Finally
we perform a combined fit using data
from $|\epsilon_K|$, $\Delta m_{B_{d,s}}$,
$|V_{ub}/V_{cb}|$, $\sin2\beta$ and rare charmless hadronic $B$ decays. The
combined analysis gives
$\gamma=67^\circ$ for the best fit value and $43^\circ \sim 87^\circ$
as the 95\% C.L. allowed range.
Several comments on other methods to determine
$\gamma$ based on SU(3) symmetry are also provided.
\end{abstract}

%
%

\newpage
\section{Introduction}

The origin of CP violation is still a mystery although it has
been observed in neutral kaon mixing more than 35 years.
One of the most promising model for CP
violation is the model proposed by Kobayashi and Maskawa in 1973\cite{1}.
This is now referred as
the Standard Model (SM) for CP violation. In this model CP violation
results from a non-removable phase $\gamma$
in the charged current mixing matrix,
the Cabbibo-Kobayashi-Maskawa (CKM) matrix\cite{1,2},
$V_{CKM}$. There are also other
mechanisms for CP violation. To understand the origin of CP violation, it is
important to study in every detail of a particular mechanism against
experimental data. In this paper we carry out a study to constrain the
CP violating phase in the SM using available experimental data.

The CKM matrix $V_{CKM}$ is a $3\times 3$ unitary matrix and is usually
written as

\begin{eqnarray}
   V_{CKM}&=&
   \left (
   \begin{array}{ccc}
         V_{ud} &V_{us} &V_{ub}  \\
         V_{cd} &V_{cs} &V_{cb}  \\
         V_{td} &V_{ts} &V_{tb}
   \end{array}
   \right ).
\end{eqnarray}
In the literature there are several ways to parameterize the CKM matrix.
The standard particle data group parameterization is given by\cite{3}

\begin{eqnarray}
V_{CKM}
   &=&
\left (
   \begin{array}{ccc}
         c_{12}c_{13} & s_{12}c_{13}& s_{13} e^{-i\delta_{13}} \\
         -s_{12}c_{23}-c_{12}s_{23}s_{13}e^{i\delta_{13}} &
c_{12}c_{23}-s_{12}s_{23}s_{13}e^{i\delta_{13}}& s_{23}c_{13}\\
         s_{12}s_{23}-c_{12}c_{23}s_{13}e^{i\delta_{13}} &
-c_{12}s_{23}-s_{12}c_{23}s_{13}e^{i\delta_{13}} & c_{23}c_{13}
   \end{array}
   \right ).
\end{eqnarray}
where $s_{ij}=\sin \theta_{ij}$ and $c_{ij}=\cos \theta_{ij}$ are the
rotation angles. A non-zero value for $\sin\delta_{13}$ violates CP.
Another commonly used parameterization is the
Wolfenstein
parameterization\cite{4} which expands the CKM matrix in terms of
$\lambda = |V_{us}|$ and is given by

\begin{eqnarray}
   V_{CKM} =
   \left (
   \begin{array}{ccc}
         1-\frac{1}{2}\lambda^2 & \lambda & A\lambda^3(\rho-i \eta) \\
          -\lambda & 1-\frac{1}{2}\lambda^2 & A\lambda^2 \\
          A\lambda^3(1-\rho-i \eta) & -A\lambda^2 &1
   \end{array}
   \right )+ {\cal O}(\lambda^4).
\end{eqnarray}
The parameters A, $\rho$, $\eta$ are of order unity.
When discussing
CP violation in kaon system, it is necessary to keep higher order terms in
$\lambda$, namely, adding $-A^2 \lambda^5 (\rho +i \eta)$ and $-A
\lambda^4 (\rho + i \eta)$ to $V_{cd}$ and
$V_{ts}$, respectively. CP violation in this parameterization is
characterized by a non-zero value for $\eta$.

Due to the unitarity condition, one has

\begin{eqnarray}
V_{ub}^*V_{ud} + V_{cb}^*V_{cd} + V_{tb}^*V_{td} = 0.
\end{eqnarray}
In the complex plane the above equation defines a triangle with
angles $\alpha=-Arg(V_{td}V_{tb}^*/V_{ud}V_{ub}^*)$,
$\beta=-Arg(V_{cd}V_{cb}^*/V_{td}V_{tb}^*)$ and
$\gamma= -Arg(V_{ud}V_{ub}^*/V_{cd}V_{cb}^*)$ as shown in Figure 1.

\begin{figure}[htb]
\centerline{ \DESepsf(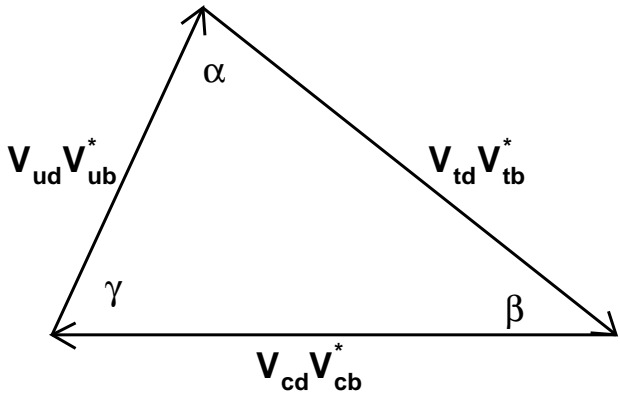 width 10cm)}
\smallskip
\caption {
The KM unitarity triangle.
} \label{triangle}
\end{figure}

To a very good approximation the phase $\delta_{13}$
is equal to $\gamma$.
In terms of $\rho$ and $\eta$, the angles $\alpha$, $\beta$ and $\gamma$ are given by:
\begin{eqnarray}
\sin 2 \alpha=\frac{2\eta ( \rho^2-\rho+ \eta^2)}{((1-\rho)^2+\eta^2)(\rho^2+\eta^2)}
\;,\qquad \sin 2 \beta=\frac{2 \eta (1-\rho)}
{(1-\rho)^2+\eta^2}\;,
\qquad \gamma=\tan^{-1} \frac{\eta}{\rho}.
\label{angle}
\end{eqnarray}

In this paper we will concentrate on obtaining constraint on
the phase $\gamma$.
Great efforts have been made to constrain or to
determine the CP violating phase $\gamma$.
Previous studies mainly used experimental data on:
i) The CP violating parameter $\epsilon_K$ in the mixing of
neutral kaons; ii) The mixing parameters $\Delta m_{B_d}$ and $\Delta m_{B_s}$
in $B_{d,s}-\bar B_{d,s}$ systems; And iii)
$|V_{ub}/V_{cb}|$ which characterizes the
strength of the charmless flavor changing and charmed flavor changing
semi-leptonic $B$ decays. The best fit value for
$\gamma$ from these considerations is around $65^\circ$\cite{5,5a}.

During the last few years, several rare
charmless hadronic $B$ decays have been measured\cite{6}.
Some of these decays are
sensitive to
$\gamma$ and therefore can be used to constrain it\cite{7,8}.
Analysis based on naive factorization approximation
suggests that $\gamma$ tends to be larger than $90^\circ$
in conflict with the analysis mentioned earlier\cite{7,8}.
If confirmed, it is
an indication of new physics beyond the SM.
Of course due to uncertainties
in the experimental data and theoretical calculations, it is not possible to
draw a firm conclusion whether this conflict is real at present.
To improve the situation,
in this paper we will carry out an analysis replacing the naive
factorization assumption by more general SU(3) flavor symmetry for
the rare
charmless hadronic
decays of $B$ to two SU(3) octet pseudoscalars $P_1$ and $P_2$,
that is, $B\to PP$ decays.

SU(3) analyses for $B$ decays have been studied by many groups and several
interesting results, such as relations between different decay branching ratios,
and ways to
constrain and/or to
determine the phase $\gamma$, have been obtained\cite{9,10,11,12,13}.
SU(3) symmetry is expected
to be a good approximation for $B$ decays. At present experimental data from
$B\to D K(\pi)$ support such an expectation\cite{9}.
However more tests are needed,
especially in rare charmless hadronic $B$ decays.
Recently it has been shown that such tests can indeed be carried out for rare
charmless hadronic $B$ decays in an electroweak model independent way in the
future\cite{12}. Before this can be done, however, SU(3) symmetry can only
be taken as a working
hypothesis. In the rest of the paper we will study constraints which can be
obtained from rare charmless hadronic $B\to PP$ decays based on SU(3) symmetry.
We will also study SU(3) breaking effects using model calculations.

The paper is arranged as follow. In section II, we will review and
up date the constraint
on $\gamma$ using information from $\epsilon_K$, $\Delta m_{B_{d,s}}$ and
$|V_{ub}/V_{cb}|$, and also information from $\sin2\beta$ measurement. 
In section III we will carry out a $\chi^2$ analysis of
$\gamma$ using rare charmless hadronic $B\to PP$ decay data based on SU(3)
symmetry. We will also discuss SU(3) breaking effects. In section IV, we
will make a combined study using results from sections II and III.
And in section V, we will discuss some of
the implications of the results obtained
and draw our conclusions.

\section{Constraint on $\gamma$ from $|\epsilon_K|$, $\Delta m_{B_{d,s}}$,
$|V_{ub}/V_{cb}|$ and $\sin2\beta$}

In this section we first review and up date
the constraint on $\gamma$ using experimental
and theoretical information on $\epsilon_K$, $\Delta m_{B_{d,s}}$ and
$|V_{ub}/V_{cb}|$. Such an analysis has been carried out before.
The analysis in this section is an up date of the previous analyses which
also serves to set up our notations for later use. We then include
experimental data from $\sin2\beta$ measurement into the analysis to obtain
the best fit value and allowed range for $\gamma$.

There exist quite a lot of information about the CKM matrix\cite{3}.
The value of $V_{us}$ is known from $K_{l3}$
decay and hyperon decays with good precision:

\begin{eqnarray*}
\lambda=0.2196 \pm 0.0023.
\end{eqnarray*}

The parameter $A$ depends on $\lambda$ and on the CKM matrix element $|V_{cb}|$.
Using experimental data from $B \to \bar D^* l^+ \nu$ and $B \to \bar D l^+ \nu$
and inclusive $b \to c l\bar \nu$, analysis from LEP data obtains
$V_{cb} = 0.0402\pm 0.0019$, and data from CLEO obtains
$V_{cb} = 0.0404\pm 0.0034$. The central values of these two measurements are
close to each other. In our analysis we will use the averaged value which
leads to
$A=0.835\pm 0.034$.

The value for $|V_{ub}|$ has also been studied using data from
$B\to \pi l\bar \nu_l$, $B\to \rho l \bar \nu_l$ and inclusive
$b\to u l\bar \nu_l$ with

\begin{eqnarray}
|V_{ub}/V_{cb}|=\lambda \sqrt{\rho^2+\eta^2} = 0.090\pm 0.025
\label{vub}
\end{eqnarray}

To separately determine $\rho$ and $\eta$ (or $\gamma$), one has to
use information from other data.
In the rest of this section we will carry out a $\chi^2$ analysis using
constraints from the measurements of $|\epsilon_K|$,
$\Delta m_{B_{d,s}}$ and $|V_{ub}/V_{cb}|$
along with other known experimental and
theoretical information.

The parameter $\epsilon_K$ indicates CP violation in neutral kaon mixing.
The short and long lived mass eigenstates $K_S$ and $K_L$
of the neutral kaons can be expressed as the linear combination of
weak interaction eigenstates $K^0$ and $\bar K^0$ as $|K_S
\rangle =p|K^0 \rangle+q|\bar K^0 \rangle$ and $|K_L \rangle
=p|K^0 \rangle -q|\bar K^0 \rangle$. $p$ and $q$ are related to
the CP violating parameter $\epsilon_K$ in $K_L$ decays by:

\begin{eqnarray}
\frac{p}{q}=\frac{1+ \epsilon_K }{1- \epsilon_K }.
\end{eqnarray}
The precise measurements of the $K_S \to \pi \pi$ and $K_L \to \pi \pi$
decay rates imply \cite{3}:

\begin{eqnarray*}
|\epsilon_K|=(2.271 \pm 0.017) \times 10^{-3}.
\end{eqnarray*}
Evaluating the so called ``Box'' diagram, one obtains

\begin{eqnarray}
|\epsilon_K|=\frac{G_F^2 f_K^2 m_K m_W^2}{6 \sqrt{2} \pi^2 \Delta m_K}
B_K (A^2 \lambda^6 \eta)\left[ y_c(\eta_{ct}f_3(y_c,y_t)-\eta_{cc})
+\eta_{tt}y_t f_2(y_t) A^2 \lambda^4 (1-\rho)
\right].
\end{eqnarray}
where
 $\eta_{tt}=0.574 \pm 0.004$,
$\eta_{ct}=0.47 \pm 0.04$ and $\eta_{cc}=1.38 \pm 0.53$\cite{14}
are the QCD correction factors,
$\Delta m_K=m_{K_L}-m_{K_S}=(0.5300 \pm 0.0012) \times 10^{-2} \mbox{ps}^{-1}$,
and $B_K=0.94 \pm 0.15$\cite{15} is the bag factor.
The functions $f_2$ and $f_3$ of the variables $y_t=m_t^2/m_W^2$ and
 $y_c=m_c^2/m_W^2$ are given by\cite{16}:

\begin{eqnarray}
f_2(x)&=& \frac{1}{4}+ \frac{9}{4(1-x)}- \frac{3}{2(1-x)^2}-\frac{3x^2 \ln x}{2(1-x)^3}\;, \nonumber\\
f_3(x,y)&=& \ln \frac{y}{x}-\frac{3y}{4(1-y)}\left( 1+\frac{y \ln y}{1-y} \right).\label{dmd}
\end{eqnarray}

Neutral mesons $B_d^0$ and $\bar B_d^0$ show a behavior similar to
neutral kaons. The heavy and light mass eigenstates, $B_L$ and
$B_H$, are different from $B_d^0$
and $\bar B_d^0$ and are given by

\begin{eqnarray}
|B_L \rangle &=&p|B_d^0 \rangle+q|\bar B_d^0 \rangle\;, \nonumber\\
|B_H \rangle &=&p|B_d^0 \rangle-q|\bar B_d^0 \rangle.
\end{eqnarray}

The mass difference $\Delta m_{B_d}=m_{B_H}-m_{B_L}$
 can be measured by means of the study of the oscillations of
one CP eigenstate into the other. The world average value for
$\Delta m_{B_d}$ is \cite{17}:

\begin{eqnarray}
\Delta m_{B_d} =0.487 \pm 0.014 \mbox{ ps}^{-1}.
\end{eqnarray}

The contribution to $\Delta m_{B_d}$ is from analogous ``Box''
diagrams as that for $\epsilon_K$, but with the dominant contribution
from the top quark in the loop. One obtains

\begin{eqnarray}
\Delta m_{B_d} =
{\frac{G_{F}^{2}}{{6\pi ^{2}}}}m_{W}^{2}m_{B_{d}}(f_{B_{d}}
\sqrt{B_{B_{d}}%
})^{2}\eta _{B}y_{t}f_{2}(y_{t})A^2\lambda ^{6}[(1-\rho )^{2}+\eta
^{2}].
\end{eqnarray}
where $f_{B_{d}}\sqrt{B_{B_{d}}}=0.215 \pm 0.040 \mbox{GeV}$\cite{18},
$\eta_B=0.55 \pm 0.01$\cite{14} and the function $f_2$ is given by
Eq. (\ref{dmd}).

$B_s^0$ and $\bar B_s^0$ mesons are believed to undergo a mixing
analogous to the $B_d^0$ and $\bar B^0_d$. Their larger mass difference
$\Delta m_{B_s}$ is responsible for oscillations that are faster than
the $B_d^0$ and $\bar B^0_d$ oscillation,
and have thus still eluded direct observation. A
lower limit has been set by the LEP, SLD and CDF collaborations, as \cite{17}:

\begin{eqnarray}
\Delta m_{B_s} > 14.9 \mbox{ps}^{-1} \mbox{(95\% C.L.)}.
\end{eqnarray}

The expression for $\Delta m_{B_s}$ in the SM is similar
to that for $\Delta m_{B_d}$. $\Delta m_{B_s}$ can be written as:

\begin{eqnarray}
\Delta m_{B_s} =\Delta m_{B_d} \frac{1}{\lambda^2} \frac{m_{B_s}}{m_{B_d}}\xi^2
\frac{1}{(1-\rho)^2+\eta^2},
\end{eqnarray}
where all the theoretical uncertainties are included in a quantity $\xi$,
which is given by\cite{18}:

\begin{eqnarray}
\xi=\frac{f_{B_s} \sqrt{B_{B_s}}}{f_{B_d} \sqrt{B_{B_d}}}=1.14\pm 0.06 .
\end{eqnarray}

The $\rho$ and $\eta$ parameters can be determined from a fit to
the experimental values of the observables described in the
above. In the analysis we will adopt the strategies used in previous analysis
in the literature
fixing the known
parameters, theoretical or experimental,
to their central values if their errors were reasonably
small reported in the left half of Table I.
The quantities affected by large errors will be used as
additional parameters of the fit, but including a constraint on
their value as shown by the right half of Table I.
All errors will be assumed to be
Gaussian. This assumption may result in stringent constraints more than
actually can be achieved because some of the errors may obey different
distributions,
for example those errors come from theoretical estimates may obey flat
distribution.
Nevertheless, the results provide a good indication for the values of the
parameters involved.

To obtain the best fit values and certain confidence level allowed ranges
for the relevant parameters, we perform a $\chi^2$ analysis using the above
information.
The procedure for $\chi^2$ analysis here is to
minimize the following expression:

\begin{eqnarray}
\chi ^{2} &=&\frac{(\widehat{A}-A)^{2}}{\sigma _{A}^{2}}+\frac{(\widehat{%
m_{c}}-m_{c})^{2}}{\sigma _{m_{c}}^{2}}+\frac{(\widehat{m_{t}}-m_{t})^{2}}{%
\sigma _{m_{t}}^{2}}+\frac{(\widehat{B_{K}}-B_{K})^{2}}{\sigma _{B_{K}}^{2}}+%
\frac{(\widehat{\eta _{cc}}-\eta _{cc})^{2}}{\sigma _{\eta
_{cc}}^{2}}
\nonumber \\
&+&\frac{(\widehat{\eta _{ct}}-\eta _{ct})^{2}}{\sigma _{\eta _{ct}}^{2}}+%
\frac{(\widehat{f_{B_{d}}\sqrt{B_{B_{d}}}}-f_{B_{d}}\sqrt{B_{B_{d}}})^{2}}{%
\sigma _{f_{B_{d}}\sqrt{B_{B_{d}}}}^{2}}+\frac{(\widehat{\xi }-\xi )^{2}}{%
\sigma _{\xi }^{2}}+\frac{(\widehat{\frac{|V_{ub}|}{|V_{cb}|}}-\frac{|V_{ub}|%
}{|V_{cb}|})^{2}}{\sigma _{\frac{|V_{ub}|}{|V_{cb}|}}^{2}}  \nonumber \\
&+&\frac{(\widehat{|\epsilon _{K}|}-|\epsilon _{K}|)^{2}}{\sigma
_{|\epsilon _{K}|}^{2}}+\frac{(\widehat{\Delta m_{B_d}}-\Delta
m_{B_d})^{2}}{\sigma _{\triangle m_{B_d}}^{2}}+\chi ^{2}(A(\Delta
m_{B_s}),\sigma _{A}(\Delta m_{B_s})).
\end{eqnarray}

The symbols with a hat represent the reference values measured or
calculated for given physical quantities, as listed in Table
I, while the corresponding $\sigma$ are their
errors. The parameters of the fit are $\rho$ , $\eta$, A, $m_c$,
$m_t$, $B_K$, $\eta_{ct}$, $\eta_{cc}$, $f_{B_d} \sqrt{B_{B_d}}$
and $\xi$.

\begin{table}[htb]
\begin{center}
\caption{Input parameters for $\chi^2$ analysis using data from
$\epsilon_K$, $\Delta m_{B_{d,s}}$ and $|V_{ub}/V_{cb}|$.}
\begin{tabular}{|l|l||l|l|}
 Fixed values & & Varied parameters& \\
\hline \hline $\lambda
=0.2196\pm 0.0023$ &\cite{3}& ${\rm A}=0.835\pm
0.034$&\cite{3} \\
$G_{F}=(1.16639\pm 0.00001)\times
10^{-5}\mbox{ GeV}^{-2}\;\;\;$ &\cite{3}& $\eta _{ct}=0.47\pm
0.04$&\cite{14}\\
$f_{K}=0.1598\pm 0.0015GeV$ &\cite{3}& $\eta _{cc}=1.38\pm
0.53$&\cite{14} \\
$\Delta m_{K}=(0.5300\pm 0.0012)\times
10^{-2}\mbox{ ps}^{-1}$ &\cite{3}& $\overline{m}_{c}%
(m_{c})=1.25\pm 0.10\mbox{ GeV}$&\cite{3} \\
$m_{K}=0.497672\pm
0.000031\mbox{ GeV}$ &\cite{3}& $\overline{m}_{t}(m_{t})=165.0\pm
5.0\mbox{ GeV}$&\cite{3}\\
$m_{W}=80.419\pm 0.056\mbox{ GeV}$
&\cite{3}& $f_{B_{d}}\sqrt{B_{B_{d}}}=0.215\pm 0.040\mbox{
GeV}\;\;\;$&\cite{15}\\
$m_{B_{d}}=5.2794\pm 0.0005\mbox{ GeV}$
&\cite{3}& $B_{K}=0.94\pm 0.15$&\cite{15}\\
 $m_{B_{s}}=5.3696\pm 0.0024\mbox{ GeV}$ &\cite{3}& $\xi
=1.14\pm 0.06$&\cite{15}\\
$\eta_{tt}=0.574\pm 0.004$ &\cite{14}&
$|\epsilon_{K}|=(2.271\pm 0.017)\times 10^{-3}$&\cite{3} \\
$\eta_{B}=0.55\pm 0.01$ &\cite{14}& $\Delta m_{B_d}=0.487\pm 0.014\mbox{
ps}^{-1}$&\cite{17}\\
&&$|V_{ub}/V_{cb}|=0.090\pm 0.025$&\cite{3}
\\ 
\end{tabular}
\end{center}
\end{table}

The inclusion of the $\Delta m_{B_s}$ data needs some explanation.
The experimental data consists of measured values of ${\cal
A}(\Delta m_{B_s})$ and $\sigma_{\cal A}(\Delta m_{B_s})$ for various
values of $\Delta m_{B_s}$ plot in Figure 2. To include this
data in the fit, for each set of free parameters $(A,\rho , \eta ,
\xi )$ we calculate the value of $\Delta m_{B_s}$ and find the
corresponding experimental values of ${\cal A}$ and $\sigma_{\cal
A}$ in Figure 2.
A nonzero value of $\Delta m_{B_s}$
implies that there is
$B_s^0-\bar B^0_s$ mixing, if observed one should have ${\cal
A}=1$ and otherwise ${\cal A}=0$\cite{bsmixing}.
We follow Ref.\cite{5a} to add to the total $\chi^2$ in Eq. (16)
a $\Delta \chi^2$ for the corresponding set of $(A,\rho,\eta,\xi)$,

\begin{equation}
\Delta \chi^2 = \chi^2({\cal A} (\Delta m_{B_s}),\sigma_{\cal A}(\Delta m_{B_s}))
=(\frac{{\cal A}-1}{\sigma_{\cal A}})^2.
\end{equation}
$Exp[-\Delta\chi^2/2]$ is an indication of how likely a mixing
with a given $\Delta m_{B_s}$ was measured by experiment. 
The sign of the deviation ${\cal A}-1$ should also be carefully treated.
Naively the expression of $\Delta \chi^2$ implies 
that a lower probability is attirubted to the $\Delta m_{B_s}$ values with
${\cal A}>1$ with respect to $\Delta m_{B_s}$ values having ${\cal A}=1$.
To avoid this undesired behavior, we follow Ref.\cite{5a} to set ${\cal A}$
to unity for the range with
$\cal A$ larger than one in Fig. 2.

\begin{figure}[htb]
\centerline{ \DESepsf(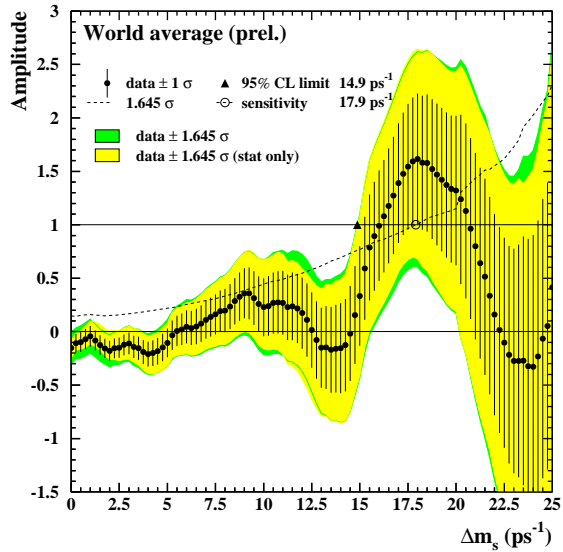 width 8cm)}
\smallskip
\caption {
Experimental data on $\Delta m_{B_s}$[17].
} \label{dms}
\end{figure}

After $\rho$ and $\eta$ are determined it
is easy to obtain the values of the angles in the unitarity
triangle using relations in Eq. (\ref{angle}).
The best fit values and the allowed regions in the
$\rho-\eta$ plane are shown in Figure 3.
The best fit values and their 68\% C.L. errors are

\begin{eqnarray}
&&\rho=0.18 ^{+0.11}_{-0.09},\;\;\; \eta=0.34^{+0.07}_{-0.06}\;,\nonumber\\
&&\sin 2 \alpha=-0.19^{+0.37}_{-0.42},\;\;\; \sin 2
\beta=0.70^{+0.14}_{-0.09},\;\;\; \gamma=62^{+12^ \circ}_{-13^
\circ}.
\label{r1}
\end{eqnarray}
The $95\%$ C.L. allowed regions for the above quantities are
expressed as:

\begin{eqnarray}
&&0.03<\rho<0.38, \;\;\;0.23<\eta<0.50, \nonumber\\
&&-0.85 <\sin
2\alpha<0.42\;,\;\;0.49 <\sin 2\beta<0.94\;,\;\;
39 ^{\circ} <\gamma<84^{\circ}.
\label{r2}
\end{eqnarray}
These results agree with previous analyses\cite{5}.

\begin{figure}[htb]
\centerline{ \DESepsf(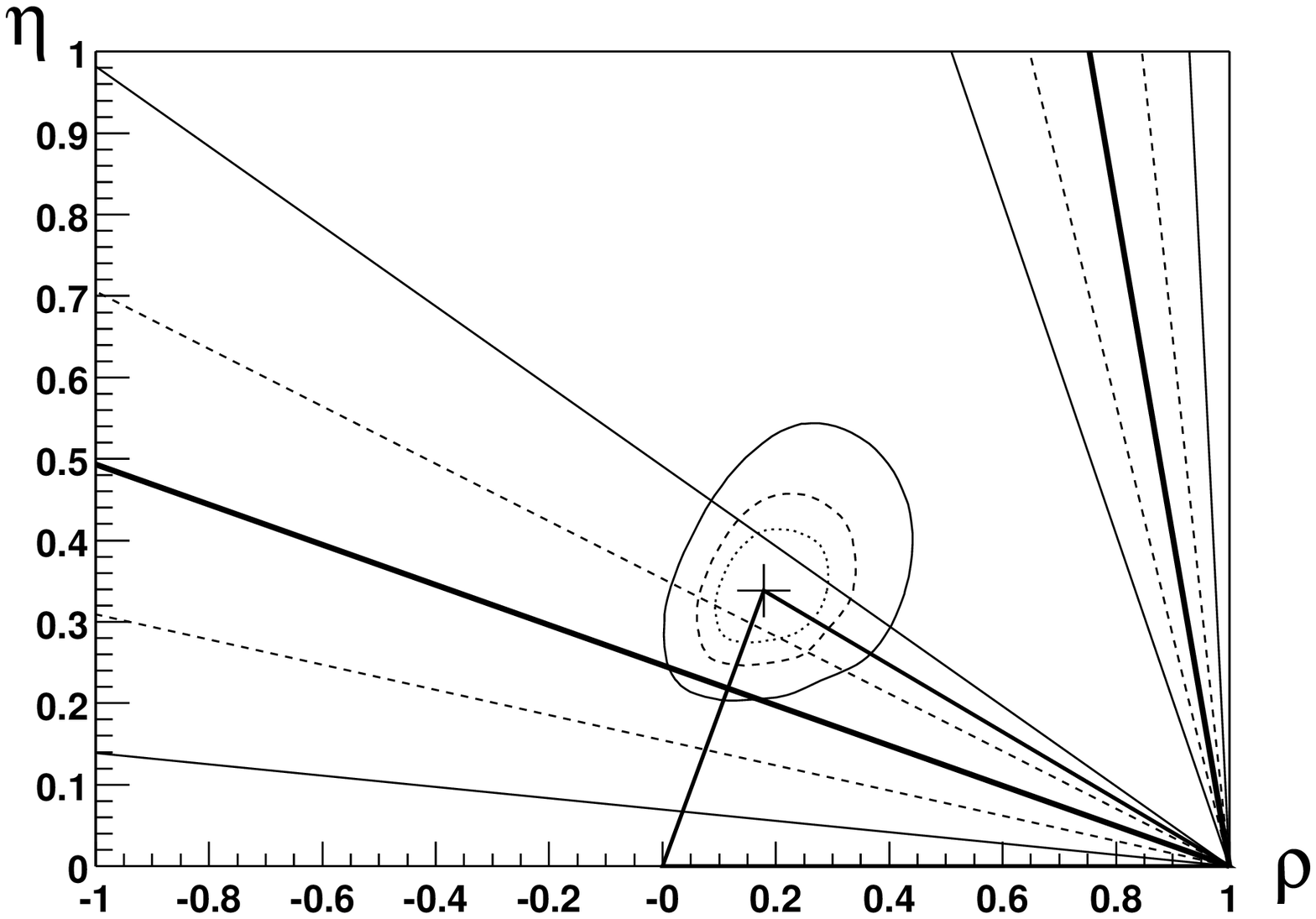 width 7cm) \DESepsf(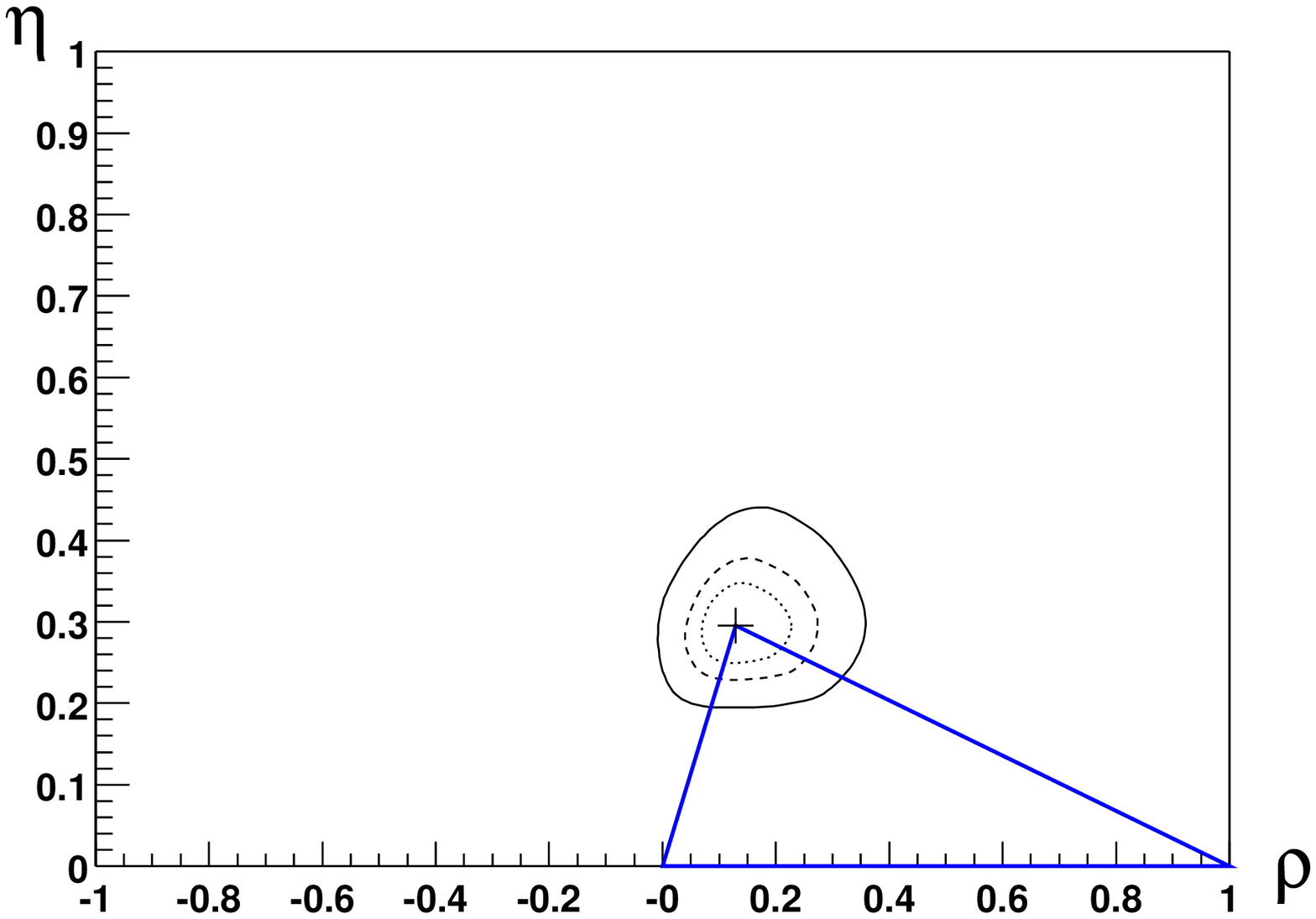 width 7cm)}

\smallskip
\caption {Constraints on $\rho$ and $\eta$ using data from
$|\epsilon_K|$, $\Delta m_{B_{d,s}}$ and $|V_{ub}/V_{cb}|$, and
$\sin2\beta$.
In the figure on the left, only $|\epsilon_K|$, $\Delta m_{B_{d,s}}$
and $|V_{ub}/V_{cb}|$ are used.  
The best fit value is indicated by the ``+'' symbol.
The region in the dotted curve corresponds to $\chi^2 - \chi^2_{min} =1$
allowed region which is at the 39\% C.L..
The 68\% C.L. allowed region is within the dashed curve and the
95\% C.L. allowed region is within the solid curve.
The straight ray lines are the results for direct measurement
of $\sin2\beta$. The thick solid lines are for the central value of
$\sin2\beta$. There are two allowed regions.
The region outside the two thin solid straight lines for each
allowed region
are excluded by the $\sin2\beta$ measurement at 95\% C.L.. The 68\% C.L.
allowed regions are between the dashed lines.
The figure on the right 
is a fit with $\sin2\beta$ data also included in the
$\chi^2$.} \label{fit}
\end{figure}

The solid line in Figure 4 is a plot of 
the minimal $\chi^2$ as a function of $\gamma$
for the fit using $|\epsilon_K|$, $\Delta m_{B_{d,s}}$ and
$|V_{ub}/V_{cb}|$. It is
clear that $\chi^2$ changes with $\gamma$ dramatically. When going away
from the minimal, $\chi^2$ raises rapidly indicating a good determination
of $\gamma$.

\begin{figure}[htb]
\centerline{ \DESepsf(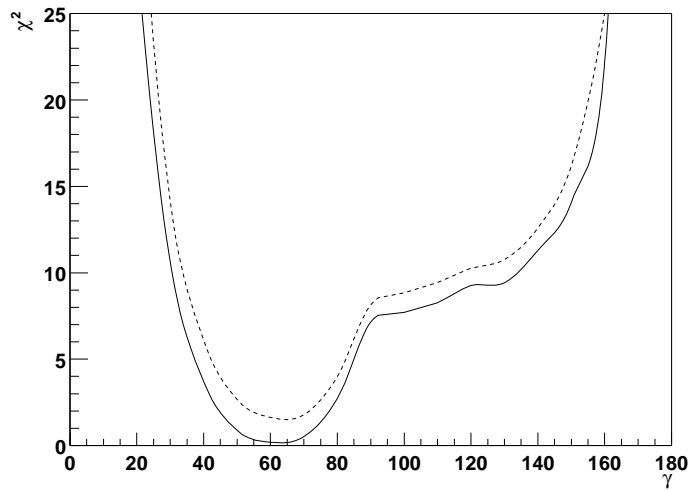 width 10cm)}
\smallskip
\caption {The solide line is the
$\chi^2$ as a function of $\gamma$ using data from $|\epsilon_K|$,
$\Delta m_{B_{d,s}}$ and $|V_{ub}/V_{cb}|$. The dashed line is the
$\chi^2$ as a function of $\gamma$
with $\sin 2\beta$ data included in the fit.
}
\end{figure}

There are also direct measurements of $\sin2\beta$ by several groups from
the time
dependent CP asymmetry in $B\to J/\psi K_S$. In the SM
this asymmetry is given by

\begin{eqnarray}
a(t) = {\Gamma(\bar B^0(t) \to J/\psi K_S) -\Gamma(B^0(t) \to J/\psi K_S)
\over
\Gamma(\bar B^0(t) \to J/\psi K_S) +\Gamma(B^0(t) \to J/\psi K_S)}
=-\sin 2\beta \sin (\Delta m_{B_d} t ).
\end{eqnarray}

The values measured by different groups are,

\begin{eqnarray}
\sin 2 \beta=\left\{ \begin{array}{ll} 0.34 \pm 0.20 \pm
0.05;  \,\,\,\,\,\,\,\,& \mbox{BaBar}\cite{19}
\\ 0.58^{+0.32+0.09}_{-0.34-0.10};& \mbox{Belle}\cite{19}\\
0.79^{+0.41}_{-0.44};& \mbox{CDF}\cite{20}\\
0.84^{+0.82}_{-1.04}\pm 0.16 &
\mbox{ALEPH}\cite{21}\end{array} \right.
\end{eqnarray}
The averaged value is  $\sin 2 \beta=0.46 \pm 0.16$.

For a given $\sin2\beta$ there are, in general, four solutions for
$\gamma$ with two of them having negative $\eta$ and another two having
positive $\eta$. To determine which one of them is the right solution,
one has to use other information. Using the information from our previous
fit, we can rule out some of the solutions.
The allowed ranges for $\rho$ and $\eta$ from the averaged value
for $\sin2\beta$ is shown in the figure on the left in 
Fig. 3 by the straight ray lines.
Since the fit from $|\epsilon_K|$, $\Delta m_{B_{d,s}}$ and $|V_{ub}/V_{cb}|$
determines $\eta >0$, only solutions with $\eta >0$ are allowed.
It is clear that one of the values for $\rho$ and $\eta$
determined from $\sin2\beta$
measurement can be
consistent with the fitting results in Eqs. (\ref{r1}) and (\ref{r2}).

It is intersting to note that $\sin2\beta$ data can eliminate a large
allowed range in the $\rho$ vs. $\eta$ plane at the 95\% C.L. level.
One can also include the measured $\sin 2 \beta$ into the $\chi^2$ analysis.
The results are shown in the figure on the right in Fig.3.
The $\chi^2$ as a function of $\gamma$ is shown by the dashed line in 
Fig. 4.
The best fit values and their 68\% C.L. errors are given by:

\begin{eqnarray}
&&\rho=0.13 ^{+0.10}_{-0.06},\;\;\; \eta=0.30^{+0.05}_{-0.05}\;,\nonumber\\
&&\sin 2 \alpha=-0.19^{+0.35}_{-0.44},\;\;\; \sin 2
\beta=0.61^{+0.09}_{-0.07},\;\;\; \gamma=66^{+10^ \circ}_{-14^
\circ}.
\end{eqnarray}

And the 95\% C.L. allowed regions for the above quantities are:

\begin{eqnarray}
&&0.01<\rho<0.30, \;\;\;0.21<\eta<0.41, \nonumber\\
&&-0.88 <\sin
2\alpha<0.45\;,\;\;0.40 <\sin 2\beta<0.80\;,\;\;
42 ^{\circ} <\gamma<87^{\circ}.
\end{eqnarray}

\section{Determination of $\gamma$ from charmless hadronic $B$ decays}

In this section we study how the phase $\gamma$ can be constrained from
experimental data on $B\to PP$ decays,
based on flavor SU(3) symmetry consideration.

\subsection{The Quark Level Effective Hamiltonian}

The quark level effective Hamiltonian up to one loop level in
electroweak interaction
for charmless hadronic $B$ decays,
including QCD corrections to the matrix elements, can be written as
\begin{eqnarray}
 H_{eff}^q = {G_{F} \over \sqrt{2}} [V_{ub}V^{*}_{uq} (c_1 O_1 +c_2 O_2)
   - \sum_{i=3}^{11}(V_{ub}V^{*}_{uq}c_{i}^{uc} +V_{tb}V_{tq}^*
   c_i^{tc})O_{i}].
\end{eqnarray}
The coefficients
$c_{1,2}$ and $c_i^{jk}=c_i^j-c_i^k$, with $j$ indicates the internal quark,
are the Wilson Coefficients (WC). These
WC's have been evaluated by several groups\cite{22},
with $|c_{1,2}|>> |c_i^j|$.
In the above the factor $V_{cb}V_{cq}^*$ has
been eliminated using the unitarity property of the CKM matrix. The
operators $O_i$ are defined as\cite{22},

\begin{eqnarray}
\begin{array}{ll}
O_1=(\bar q_i u_j)_{V-A}(\bar u_j b_i)_{V-A}\;, &
O_2=(\bar q u)_{V-A}(\bar u b)_{V-A}\;,\\
O_{3,5}=(\bar q b)_{V-A} \sum _{q'} (\bar q' q')_{V \mp A}\;,&
O_{4,6}=(\bar q_i b_j)_{V-A} \sum _{q'} (\bar q'_j q'_i)_{V \mp A}\;,\\
O_{7,9}={ 3 \over 2} (\bar q b)_{V-A} \sum _{q'} e_{q'} (\bar q' q')_{V \pm A}\;,\hspace{0.3in} &
O_{8,10}={ 3 \over 2} (\bar q_i b_j)_{V-A} \sum _{q'}
e_{q'} (\bar q'_j q'_i)_{V \pm A}\;,\\
O_{11} = {g_s\over 8 \pi^2} \bar q \sigma_{\mu\nu} G^{\mu\nu} (1+\gamma_5) b\;.&
\end{array}
\end{eqnarray}
where $(\bar q_1 q_2)_{V-A} = \bar q_1 \gamma_\mu (1-\gamma_5) q_2$,
$G^{\mu\nu}$ is the field strengths of the gluon, respectively.
We have neglected the photonic dipole penguin term whose contribution to
hadronic charmless $B$ decays is small.
The usual tree-level W-exchange
contribution in the effective Hamiltonian corresponds to $O_2$.
$O_1$ emerges due to the QCD
corrections. The operators $O_{3,4,5,6}$ are from the
QCD-penguin diagrams.
The operators $O_7$,..., $O_{10}$ arise from the electroweak-penguin
diagrams. $O_{11}$ is the gluonic dipole penguin operator.

The  WC's at $\mu=5$ GeV with $\alpha_s(m_Z)=0.118$,
in the regularization independent scheme in Ref.\cite{23} are

\begin{eqnarray}
\begin{array}{ll}
c_1=-0.313\;,&
c_2=1.150\;, \\
c_{3}^t=0.017\;,&
c_{4}^t=-0.037\;,\\
c_{5}^t=0.010\;, &
c_{6}^t=-0.046\;,\\
c_{7}^t=-0.001 \alpha_{em}\;,\hspace{0.3in}&
c_{8}^t=0.049 \alpha_{em}\;,\\
c_{9}^t=-1.321 \alpha_{em}\;,&
c_{10}^t=0.267 \alpha_{em}\;\\
c_{11}^t = -0.143\;.&
\end{array}
\end{eqnarray}
where $\alpha_{em}=1/128$. $c^{c,u}_i$ are given in Ref.\cite{23}

\subsection{SU(3) Structure of the Effective Hamiltonian}

To obtain $B$ decay amplitudes, one has to calculate hadronic matrix
elements from quark operators. At present there is no reliable
methods to calculate these matrix elements although simple factorization
calculations provide some reasonable results for some decays, but not all
of them\cite{24}.
It motivates us to
carry out model independent analysis by studying properties
of the effective Hamiltonian under SU(3) flavor symmetry and use them
to obtain information about related decays.

 In general the decay amplitudes for $B\to PP$ can be written as

\begin{eqnarray}
A(B\to PP) = <P\;P|H_{eff}^q|B> = {G_F\over \sqrt{2}}[V_{ub}V^*_{uq} T(q)
+ V_{tb}V^*_{tq}P(q)]\;,
\label{aa}
\end{eqnarray}
where $T(q)$ contains contributions from
the $tree$ operators $O_{1,2}$ as well as $penguin$ operators $O_{3-11}$
due to charm and up
quark loop corrections to the matrix elements,
while $P(q)$ contains contributions purely from
$penguin$ due to top and charm quarks in loops. The amplitude $T$
in Eqs. (\ref{aa}) is usually called the ``tree'' amplitude which will also
be referred to later on in the paper. One should, however,
keep in mind that
it contains the usual tree
current-current contributions proportional to $c_{1,2}$ and also the
u and c penguin contributions proportional to $c_i^{uc}$ with
$i=3-11$.
Also, in general, it contains long distance contributions corresponding
to internal u and c generated intermediate hadron states.
In our later analysis, we do not distinguish between the tree and the penguin
contributions in the amplitude $T$.

The relative strength of the amplitudes
$T$ and $P$ is predominantly determined by their corresponding WC's in the
effective Hamiltonian.
For $\Delta S = 0$ charmless decays, the dominant contributions are due to the
tree operators $O_{1,2}$ and the penguin operators are suppressed by smaller
WC's. Whereas for $\Delta S =-1$ decays, because the penguin contributions are
enhanced by a factor of $V_{tb} V_{ts}^*/V_{ub}V_{us}^*\approx 50$\cite{3}
compared with the
tree contributions, penguin effects dominate the
decay amplitudes. In this case the electroweak penguins
can also play a very important role\cite{25}.

The operators $O_{1,2}$, $O_{3-6,11}$, and $O_{7-10}$ transform under
SU(3) symmetry as $\bar 3 + \bar 3' +6 + \overline {15}$, $\bar
3$, and $\bar 3 + \bar 3' +6 + \overline {15}$, respectively.
We now give details for the decomposition under SU(3) for some operators.
For $\Delta S=0$ decays, $O_2$ can be
written, omitting the Lorentz-Dirac structure, as\cite{12}:
\begin{eqnarray}
O_2&=& -\frac{1}{8} \{(\bar u u)(\bar d b)
+(\bar d d)(\bar d b)
+(\bar s s)(\bar d b)\}_{\bar 3}\nonumber\\
&+&\frac{3}{8} \{(\bar d u)(\bar u b)
+(\bar d d)(\bar d b)
+(\bar d s)(\bar s b)\}_{\bar 3 '}\nonumber\\
&-&\frac{1}{4}\{ (\bar u u)(\bar d b)-(\bar d u)(\bar u b) +
(\bar d s)(\bar s b)-(\bar s s)(\bar d b)
\}_6\nonumber\\
&+&\frac{1}{8} \{3(\bar u u)(\bar d b)
+3(\bar d u)(\bar u b)
-(\bar d s)(\bar s b)-(\bar s s)(\bar d b)
-2(\bar d d)(\bar d b)\}_{\overline {15}}\nonumber\\
&=&-\frac{1}{8} H(\bar 3)+\frac{3}{8} H(\bar 3')-\frac{1}{4}H(6)
+\frac{1}{8}H(\overline{15})
\end{eqnarray}
The $\bar 3$, 6 and $\overline{15}$ indicate the SU(3) irreducible representations.
The non-zero entries of the matrices H(i) in flavor space are\cite{9}:
\begin{eqnarray}
&&H(\bar 3)^2 = H(\bar 3')^2 = 1\;,\;\;
H(6)^{12}_1 = H(6)^{23}_3 = 1\;,\;\;H(6)^{21}_1 = H(6)^{32}_3 =
-1\;,\nonumber\\
&&H(\overline{15} )^{12}_1 = H(\overline{15} )^{21}_1 = 3\;,\;
H(\overline{15} )^{22}_2 =
-2\;,\;
H(\overline{15} )^{32}_3 = H(\overline{15} )^{23}_3 = -1\;.
\end{eqnarray}
Here $1=u$, $2=d$ and $3=s$ with the upper indices indicating anti-quarks and
the lower ones indicating quarks.

For $\Delta S=1$ decays, one has
\begin{eqnarray}
O_2&=& -\frac{1}{8} \{(\bar u u)(\bar s b)
+(\bar d d)(\bar s b)
+(\bar s s)(\bar s b)\}_{\bar 3}\nonumber\\
&+&\frac{3}{8} \{(\bar s u)(\bar u b)
+(\bar s d)(\bar d b)
+(\bar s s)(\bar s b)\}_{\bar 3'}\nonumber\\
&-&\frac{1}{4}\{ (\bar u u)(\bar s b) -(\bar s u)(\bar u b)  +
(\bar s d)(\bar d b) -(\bar d d)(\bar s b) \}_6\nonumber\\
&+&\frac{1}{8} \{3(\bar u u)(\bar s b)
+3(\bar s u)(\bar u b)
-(\bar s s)(\bar s b)-(\bar s d)(\bar d b)
-2(\bar d d)(\bar s b)\}_{\overline{15}}\nonumber\\
&=&-\frac{1}{8} H(\bar 3)+\frac{3}{8} H(\bar 3')-\frac{1}{4}H(6)
+\frac{1}{8}H(\overline{15})
\end{eqnarray}
The non-zero entries are\cite{9}:
\begin{eqnarray}
&&H(\bar 3)^3 =H(\bar 3')^3 = 1\;,\;\;
H(6)^{13}_1 = H(6)^{32}_2 = 1\;,\;\;H(6)^{31}_1 = H(6)^{23}_2 =
-1\;,\nonumber\\
&&H(\overline{15} )^{13}_1 = H(\overline{15} ) ^{31}_1 = 3\;,\;
H(\overline{15} )^{33}_3 =
-2\;,\;
H(\overline{15} )^{32}_2 = H(\overline{15} )^{23}_2 = -1\;.
\end{eqnarray}

For $\Delta S=0$, the operators $O_{1,2}$, $O_{3-6}$, and $O_{7-10}$ can be
decomposed as

\begin{eqnarray}
O_1&=&\frac{3}{8} {\cal O}_{\bar 3}-\frac{1}{8} {\cal O}_{\bar 3 '}
+\frac{1}{4}{\cal O}_6+\frac{1}{8} {\cal O}_{\overline{15}}\;, \nonumber\\
O_2&=&-\frac{1}{8} {\cal O}_{\bar 3}+\frac{3}{8} {\cal O}_{\bar 3 '}
-\frac{1}{4}{\cal O}_6+\frac{1}{8} {\cal O}_{\overline{15}}\;,\nonumber\\
O_3&=&{\cal O}_{\bar 3}\;, \;\;\;\;O_4={\cal O}_{\bar 3'}\;,\nonumber\\
O_9&=&\frac{3}{2}O_1-\frac{1}{2}O_3\;, \;\;\;\;
O_{10}=\frac{3}{2}O_2-\frac{1}{2}O_4.
\label{su3d}
\end{eqnarray}
where
\begin{eqnarray}
{\cal O}_{\bar 3}&=&(\bar u u)_{V-A}(\bar d b)_{V-A}
+(\bar d d)_{V-A}(\bar d b)_{V-A}
+(\bar s s)_{V-A}(\bar d b)_{V-A}\;,\nonumber\\
{\cal O}_{\bar 3'}&=&(\bar d u)_{V-A}(\bar u b)_{V-A}
+(\bar d d)_{V-A}(\bar d b)_{V-A}
+(\bar d s)_{V-A}(\bar s b)_{V-A}\;,\nonumber\\
{\cal O}_{6}&=&(\bar u u)_{V-A}(\bar d b)_{V-A}
-(\bar d u)_{V-A}(\bar u b)_{V-A}\nonumber\\
&+&(\bar d s)_{V-A}(\bar s b)_{V-A}
-(\bar s s)_{V-A}(\bar d b)_{V-A}\;,\nonumber\\
{\cal O}_{\overline{15}}&=&3(\bar u u)_{V-A}(\bar d b)_{V-A}
+3(\bar d u)_{V-A}(\bar u b)_{V-A}
-(\bar d s)_{V-A}(\bar s b)_{V-A}\nonumber\\
&-&(\bar s s)_{V-A}(\bar d b)_{V-A}
-2(\bar d d)_{V-A}(\bar d b)_{V-A}.
\end{eqnarray}

The operators $O_5$ and $O_6$ have same SU(3) structure as $O_3$ and $O_4$
but different Lorentz-Dirac structures.
$O_7$, $O_8$ have the same SU(3) structure as
$O_9$, $O_{10}$, but again have different Lorentz-Dirac structures.
Similarly one can obtain the
decomposition  of the operators for $\Delta S=1$ case.

Since we are only concerned with flavor structure in SU(3), operators with
different Lorentz-Dirac
structures and different color structures can be grouped
together according to their flavor SU(3) representations without affect the
results. As long as flavor structure is concerned, the effective Hamiltonian
contains only $\bar 3$, $6$ and $\overline{15}$.
These properties enable us to
write the decay amplitudes for $B\to P P$ in only a few SU(3)
invariant amplitudes.

\subsection{SU(3) Decay Amplitudes for $B\to PP$ Decays}

We will use $B_i = (B_u,  B_d,  B_s) = (B^-, \bar B^0, \bar
B^0_s)$ to indicate the SU(3) triplet for the three $B$-mesons,
and $M$ to indicate  the pseudoscalar octet $M$ which contains one
of the $P$ in the final state with

\begin{eqnarray}
M&=&\left(
  \begin{array}{ccc}
        \frac{\pi^0}{\sqrt{2}}+ \frac{\eta_8}{\sqrt{6}} &\pi^+& K^+  \\
         \pi^- &-\frac{\pi^0}{\sqrt{2}}+ \frac{\eta_8}{\sqrt{6}}  &K^0  \\
         K^- & \bar K^0 &    -2 \frac{\eta_8}{\sqrt{6}}
   \end{array} \right)\;.
\end{eqnarray}

One can write the
$T$ amplitude for $B \to PP$ as\cite{9}
\begin{eqnarray}
T&=& A_{\bar 3}^TB_i H(\bar 3)^i (M_l^k M_k^l) + C^T_{\bar 3}
B_i M^i_kM^k_jH(\bar 3)^j \nonumber\\
&+& A^T_{6}B_i H(6)^{ij}_k M^l_jM^k_l + C^T_{6}B_iM^i_jH(6
)^{jk}_lM^l_k\nonumber\\
&+&A^T_{\overline{15}}B_i H(\overline{15})^{ij}_k M^l_jM^k_l +
C^T_{\overline{15}}B_iM^i_j
H(\overline{15} )^{jk}_lM^l_k\;,
\label{tpp}
\end{eqnarray}
due to the anti-symmetric nature in exchanging the upper two indices of
$H^{ij}_k(6)$, and the symmetric structure of the two mesons in the
final states,
$C_6-A_6$ always appear together\cite{9}. We will just use $C_6$ to indicate
this combination.
There are 5 complex independent SU(3) invariant
amplitudes. The results for each individual $B$ decay mode are
shown in Table II. Similarly one can write down the expressions for the
penguin induced decay amplitudes $P$.

\begin{table}[htb]
\caption{SU(3) decay amplitudes for $B\to PP$ decays.  \label{t1} }
\footnotesize
\begin{eqnarray}
\begin{array}{l}
\hspace{-3mm}
\left.
\begin{array}{l}
\Delta S = 0\\
T^{B_u}_{\pi^-\pi^0}(d) = {8\over \sqrt{2}}C^T_{\overline{15}},\\
T^{B_u}_{\pi^- \eta_8}(d)={2\over \sqrt{6}}
(C^T_{\bar 3}  - C^T_6 + 3 A^T_{\overline {15}} + 3C^T_{\overline {15}}),\\
T^{B_u}_{K^- K^0}(d)=
C^T_{\bar 3}  - C^T_6 + 3 A^T_{\overline {15}} -C^T_{\overline {15}},\\
T^{B_d}_{\pi^+\pi^-}(d) = 2A^T_{\bar 3} +C^T_{\bar 3}
+ C^T_{6} + A^T_{\overline {15} } + 3 C^T_{\overline {15}},\\
T^{B_d}_{\pi^0\pi^0}(d)= {1\over \sqrt{2}} (2A^T_{\bar 3} +C^T_{\bar 3}
+ C^T_{6} + A^T_{\overline {15} } -5 C^T_{\overline {15}}),\\
T^{B_d}_{K^- K^+}(d)= 2(A^T_{\bar 3}  +  A^T_{\overline {15}}),\\
T^{B_d}_{\bar K^0 K^0}(d)= 2A^T_{\bar 3} +
C^T_{\bar 3} - C^T_6 - 3 A^T_{\overline {15}} - C^T_{\overline {15}},\\
T^{B_d}_{\pi^0 \eta_8}(d)= {1\over \sqrt{3}}
(-C^T_{\bar 3}  + C^T_6 + 5 A^T_{\overline {15}} + C^T_{\overline {15}}),\\
T^{B_d}_{\eta_8 \eta_8}(d)={1\over \sqrt{2}}
(2A^T_{\bar 3} + {1\over 3} C^T_{\bar 3} - C^T_6
-A^T_{\overline {15}} + C^T_{\overline {15}}),\\
T^{B_s}_{ K^+ \pi^-}(d) =
C^T_{\bar 3} + C^T_6 -  A^T_{\overline {15}} +3 C^T_{\overline {15}},\\
T^{B_s}_{ K^0 \pi^0}(d) =
-{1\over \sqrt{2}}(C^T_{\bar 3} + C^T_6 -  A^T_{\overline {15}}
-5 C^T_{\overline {15}}),\\
T^{B_s}_{K^0 \eta_8}(d)=
-{1\over \sqrt{6}}(C^T_{\bar 3}  + C^T_6 -  A^T_{\overline {15}}
-5 C^T_{\overline {15}}),
\end{array}
\right.
\left.
\begin{array}{l}
\Delta S = -1\\
T^{B_u}_{\pi^-\bar K^0}(s)= C^T_{\bar 3}
 - C^T_{6} + 3A^T_{\overline {15}} -  C^T_{\overline {15}
 },\\
T^{B_u}_{\pi^0K^-}(s)= {1\over \sqrt{2}} (C^T_{\bar 3}
  - C^T_{6} + 3A^T_{\overline {15} } +7 C^T_{\overline {15}
  })\;,\\
T^{B_u}_{\eta_8K^-}(s)= {1\over\sqrt{6}}(-C^T_{\bar 3}
   + C^T_{6} - 3A^T_{\overline {15}} +9 C^T_{\overline {15}
   }),\\
T^{B_d}_{\pi^+ K^-}(s) =  C^T_{\bar 3}
+ C^T_{6} - A^T_{\overline {15}} + 3 C^T_{\overline {15}
},\\
T^{B_d}_{\pi^0\bar K^0}(s)= -{1\over \sqrt{2}} (C^T_{\bar 3}
+ C^T_{6} - A^T_{\overline {15} } -5 C^T_{\overline {15} }),\\
T^{B_d}_{\eta_8 \bar K^0}(s)=  -{1\over \sqrt{6}} (C^T_{\bar 3}
+ C^T_{6} - A^T_{\overline {15} } -5 C^T_{\overline {15} }),\\
T^{B_s}_{\pi^+\pi^-}(s) = 2(A^T_{\bar 3}
+ A^T_{\overline {15}}),\\
T^{B_s}_{\pi^0\pi^0}(s) = \sqrt{2}(A^T_{\bar 3}
+ A^T_{\overline {15}}),\\
T^{B_s}_{K^+K^-}(s)= 2A^T_{\bar 3} +C^T_{\bar 3}
+ C^T_{6} + A^T_{\overline {15} } + 3 C^T_{\overline {15}
},\\
T^{B_s}_{K^0\bar K^0}(s)= 2A^T_{\bar 3} +C^T_{\bar 3}
- C^T_{6} -3 A^T_{\overline {15} } - C^T_{\overline {15}
},\\
T^{B_s}_{\pi^0\eta_8}(s)= {2\over \sqrt{3}}
( C^T_{6}
+2 A^T_{\overline {15}} - 2C^T_{\overline {15}
}),\\
T^{B_s}_{\eta_8\eta_8}(s)= \sqrt{2}(A^T_{\bar 3} +{2\over 3} C^T_{\bar 3}
- A^T_{\overline {15} } - 2C^T_{\overline {15}}).
\end{array}
\right.
\end{array}
\nonumber
\end{eqnarray}
\end{table}

Since there are both tree and penguin amplitudes $C^T_i$, $A^T_i$ and
$C^P_i$, $A^P_i$, there in general,
10 complex hadronic parameters (20 real parameters).
However simplications can be made by noticing that $c_{7,8}$ are very small
compared with other Wilson Coefficients, their contributions can be neglected to a very good
precision. In that case, from Eq. (\ref{su3d}), we obtain

\begin{eqnarray}
C^P_6(A^P_6) &=&
- {3\over 2}
{c_9^{tc} - c_{10}^{tc}\over
c_1-c_2-3(c_9^{uc}-c_{10}^{uc})/2}
C^T_6 (A^T_6)
\approx
- {3\over 2}
{c_9^{t}-c_{10}^t\over c_1-c_2}
C^T_6 (A^T_6)
\;,\nonumber\\
C^P_{\overline {15}}(A^P_{\overline {15}})
&=& -{3\over 2}
{c_9^{tc}+c_{10}^{tc}\over c_1+c_2-3(c^{uc}_9+c_{10}^{uc})/2}
C^T_{\overline {15}} (A^T_{\overline {15}})
\approx
-{3\over 2}
{c_9^{t}+c_{10}^{t}\over c_1+c_2}
C^T_{\overline {15}} (A^T_{\overline {15}})
.
\label{relation}
\end{eqnarray}
We have checked that the
approximation signs in the above are good to $10^{-4}$.

At leading order QCD correction, the above relations are renormalization
scale independent, and therefore to this order the coefficients
$C_i$ and $A_i$ are also so.
This can be seen from the fact that
when keeping terms which mix only between $O_{1} (O_{9})$ and
$O_2 (O_{10})$, the dominant QCD correction gives:
$c_{1(9)}(\mu) + c_{2(10)}(\mu)=
\eta^{2/\beta}[c_{1(9)}(m_W) + c_{2(10)}(m_W)]$
and
$c_{1(9)}(\mu) - c_{2(10)(\mu)}=
\eta^{-4/\beta}[c_{1(9)}(m_W) - c_{2(10)}(m_W)]$. Here
$c_{1,2,9,10}(m_W)$ are the initial values for the WC's at
the W mass scale with $c_{1(10)}(m_W) =0$,
$\eta = \alpha_s(m_W)/\alpha_s(\mu)$ and $\beta = 11-2f/3$
(f is the number of quark flavors with mass smaller than $\mu$).
These relations lead to $(c_9(\mu)\pm c_{10}(\mu))/(c_1(\mu)
\pm c_2(\mu)) = \pm c_9(m_W)
/c_2(m_W)$ independent of $\mu$.
Mixings with other operators and higher order corrections introduce
dependence on renormalization schemes.
We have checked with different renormalization schemes and find that numerically
the changes are less than 15\% for different schemes.
Although the changes are not sizable,
there are scheme dependence.
The total
decay amplitudes are not renormalization scheme dependent, therefore the hadronic
matrix elements determined depend on the renormalization scheme used to
determine the ratios, $(c_9\pm c_{10})/(c_1\pm c_{2})$. One should consistently
use the same scheme.

Using relations in Eq.(\ref{relation}), one finds that
there are less independent parameters which
we choose them to be,
$C_{\bar 3}^{T,P}(A^{T,P}_{\bar 3})$, $C_6^T$, and $C^T_{\overline {15}}
(A^T_{\overline {15}})$. Using the fact that an overall phase can be removed
without loss of generality, we will set $C^P_{\bar 3}$ to be real. There
are in fact only 13 real independent parameters for $B\to PP$ in the
SM.

One can further reduce the parameters with some dynamic considerations.
To this end we note that
the amplitudes $A_i$ correspond to annihilation contributions,
as can be seen from Eq.(\ref{tpp}) where $B_i$ mesons are contracted with
one of the index in $H(j)$, are small compared with
the amplitudes $C_i$ from model calculations and are often neglected in
factorization calculations\cite{7,24}. Neglecting
all annihilation contributions, we then have just 7 independent hadronic
parameters in the
amplitudes

\begin{eqnarray}
C_{\bar 3}^P,\;\;C_{\bar 3}^T e^{i\delta_{\bar 3}},\;\;
C^{T}_6e^{i\delta_6},\;\;
C^{T}_{\overline{15}}e^{i\delta_{\overline{15}}}.
\end{eqnarray}
The phases in the above are defined in such a way that all $C_i^{T,P}$
are real positive numbers.

We will make
the assumption that annihilation amplitudes are negligiblly small in
our later analysis and leave the verification of this assumption for future
experimental data. We point out that
this assumption can be tested using
$B_d \to K^- K^+$, $B_s \to \pi^+ \pi^-$,
$\pi^0 \pi^0$ in $B \to PP$ decays,
because these decays have only annihilation contributions as can be seen from
Table 2\cite{11,12}.

\subsection{Constraint on $\gamma$ from $B\to PP$ Decays}

We are now ready to carry out a $\chi^2$ analysis using data from
$B\to \pi\pi$ and $B\to K \pi$. The experimental data to be used are
shown in Table III.

\begin{table}
\caption{Rare hadronic charmless $B\to \pi\pi$ and $B\to K \pi$ data. The
branching ratios are in unit of $10^{-6}$.}
\begin{tabular}{|l|l|l|}
 Br and $A_{CP}$ & Data&Value used from combined data \\ \hline \hline
$Br(B\to \pi^+\pi^-)$&$4.3^{+1.6}_{-1.4}\pm 0.5$\cite{6,26}
&$4.4\pm0.9$\\
&$5.9^{+2.4}_{-2.1}\pm 0.5$\cite{27}&\\
&$4.1 \pm1.0 \pm0.7$\cite{28}&\\
$Br(B\to \pi^-\pi^0)$&$5.6^{+2.6}_{-2.3}\pm 1.7$\cite{6,26}
&$6.2\pm2.4$\\
&$7.1^{+3.6+0.9}_{-3.0-1.2}$\cite{27}&\\
$Br(B\to K^+\pi^-)$&$17.2^{+2.5}_{-2.4}\pm 1.2$\cite{6,26}&
$17.3 \pm 1.6$\\
&$18.7^{+3.3}_{-3.0}\pm 1.6$\cite{27}&\\
&$16.7\pm 1.6^{+1.2}_{-1.7}$\cite{28}&\\
$Br(B\to K^-\pi^0)$&$11.6^{+3.0+1.4}_{-2.7-1.3}$\cite{6,26}&
$13.7\pm 2.6$\\
&$17.0^{+3.7+2.0}_{-3.0-2.2}$\cite{27}&\\
$Br(B\to \bar K^0 \pi^-)$&$18.2^{+4.6}_{-4.0}\pm 1.6$\cite{6,26}
&$16.2\pm3.8$\\
&$13.1^{+5.5}_{-4.6} \pm 2.6$\cite{27}&\\
$Br(B\to K^0\pi^0)$&$14.6^{+5.9+2.4}_{-5.1-3.3}$\cite{6,26}
&$14.6\pm4.6$\\
&$14.6^{+6.1}_{-5.1}\pm 2.7$\cite{27}&\\
$Br(B\to K^-K^0)$&$<5.1(90\% \mbox{C.L.})$\cite{6,26}
&$0.6\pm1.9$\\
&$<5.0(90\% \mbox{C.L.})$\cite{27}&\\
$Br(B\to\pi^0\pi^0)$&$2.1^{+1.7+0.7}_{-1.3-0.6}$\cite{26}
&$2.1\pm1.8$\\
$A_{CP}(B\to K^-\pi^0)$&$-0.29\pm0.23$\cite{29}
&$-0.13\pm0.16$\\
&$0.019^{+0.219}_{-0.191}$\cite{27}&\\
$A_{CP}(B\to K^+\pi^-)$& $-0.04\pm0.16$\cite{29}
&$-0.003\pm0.12$\\
&$0.043\pm0.175\pm0.021$\cite{27}&\\
$A_{CP}(B\to \bar K ^0\pi^-)$&$ 0.18\pm0.24$\cite{29}
&$0.18\pm0.24$\\
\end{tabular}
\end{table}

In general the errors for the experimental data in Table III are correlated.
Due to the lack of knowledge of the error correlation from experiments, in
our analysis, for simplicity, we take them to be uncorrelated and
assume
the errors obey Gaussian distribution
taking the larger one between $\sigma_+$
and $\sigma_-$ to be on the conservative side. When combining from different
measurements, we take the weighted average.
For the data which only presented as upper bounds,
we assume them to obey Gaussian distribution and taking the
error $\sigma$ accordingly.

The $\chi^2$ analysis in this case is to minimize the $\chi^2$ given
in the below

\begin{eqnarray}
\chi^2 = \sum_i {(\hat Br(i) - Br(i))^2\over \sigma^2_{Br}(i)}
+ \sum_i {(\hat A_{CP}(i) - A_{CP}(i))^2\over \sigma^2_{CP}(i)}
+\chi^2(A, |V_{ub}/V_{cb}|),
\end{eqnarray}
where the summation on $i$ is for the available decay branching ratios and
CP asymmetries listed in
Table III.
$\sigma_{Br, CP}$ are the corresponding errors.
Here $\chi^2(A,
|V_{ub}/V_{cb}|)$ is the $\chi^2$ due to uncertainties
in $A$ and $|V_{ub}/V_{cb}|$ as in section II.
The branching ratios $Br(i)$ and CP asymmetries $A_{CP}(i)$
expressed in terms of decay amplitude
$A(i) = (G_F/\sqrt{2})[V_{ub}V_{uq}^*T(i)+ V_{tb}V_{tq}^* P(i)]$
for a particular
$B\to P_1 P_2$ are given by

\begin{eqnarray}
&&Br(i) ={1\over 32\pi m_B \Gamma_B}
(|A(i)|^2+|\bar A(i)|^2)\lambda_{P_1P_2}\;,\nonumber\\
&&A_{CP}(i) ={|A(i)|^2-|\bar A(i)|^2 \over |A(i)|^2 +|\bar A(i)|^2},
\end{eqnarray}
where $\lambda_{ij} = [(1-(m_i+m_j)^2/m_B^2)(1-(m_i-m_j)^2/m_B^2)]^{1/2}$.
The amplitudes $T(i)$ and $P(i)$ for each individual decay can be read
off from Table II.

We use
$V_{cb} =0.0402\pm 0.0019$ and $|V_{ub}/V_{cb}| =0.090\pm 0.025$ in the fitting.
The results with exact SU(3) symmetry are shown in Figures
5 and 6 by the solid curves. The best fit values for the
hadronic parameters are

\begin{eqnarray}
&&C^P_{\bar 3} = 0.13,\;\;C^T_{\bar 3} = 0.34,\;\;
C_6^T = 0.13,\;\;C^T_{\overline {15}} = 0.16,\nonumber\\
&&\delta_{\bar 3} = -27^\circ,\;\; \delta_{6} = -20^\circ,\;\;
\delta_{\overline {15}} = 35^\circ.
\end{eqnarray}

\begin{figure}[htb]
\centerline{ \DESepsf(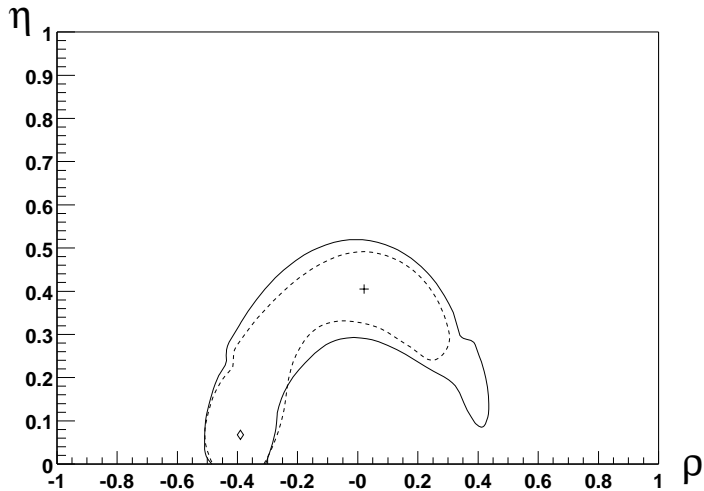 width 10cm)}
\smallskip
\caption {
The constraints for $\rho$ and $\eta$ using data from
$|V_{ub}/V_{cb}|$ and rare
$B\to \pi\pi$ and $B\to K \pi$. For the fit with exact SU(3),
the best fit value is
indicated by the ``+'' symbol and the $\chi^2-\chi^2_{min} = 1$ (39\% C.L.)
allowed regions are inside the region in the solid curve.
For the case with SU(3) breaking effects, the best fit value is indicated by
a diamond shaped symbol and the  39\% C.L. allowed region is inside the
dashed curve.
}
\end{figure}

\begin{figure}[htb]
\centerline{ \DESepsf(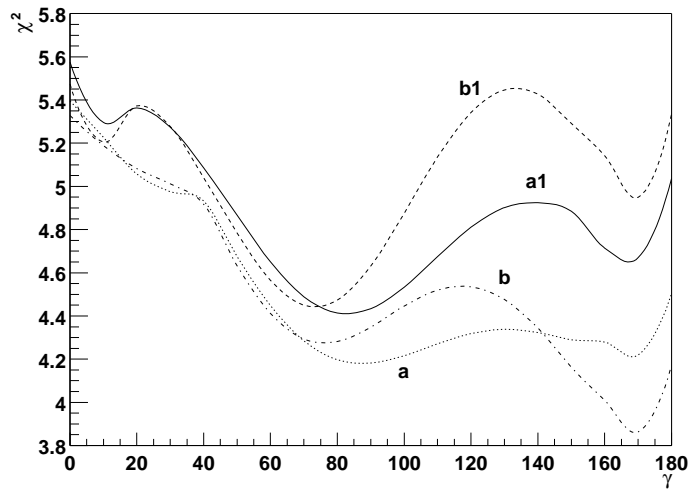 width 10cm)}
\smallskip
\caption{
$\chi^2$ as a function of $\gamma$ using data from $|V_{ub}/V_{cb}|$ and
rare
$B\to \pi\pi$ and $B\to K \pi$. The curve (a) is for the case
with exact
SU(3), and the curve (b) is for the one with
SU(3) breaking effects. The curves (a1) and (b1) are for the cases with
the additional condition $C^T_{\bar 3}e^{i\delta_{\bar 3}}
- C^T_6e^{i\delta_6} -
C^T_{\overline {15}}e^{i\delta_{\overline{15}}} = 0$
with exact SU(3) and with SU(3) breaking effects, respectively.
}
\end{figure}

And the best fit values for $\rho$, $\eta$ and $\gamma$ are

\begin{eqnarray}
\rho =0.02,\;\;
\eta = 0.40,\;\;
\gamma= 87^\circ.
\end{eqnarray}

The constraint is weak. We have not given the 68\% allowed ranges because
to that level, the constraints are basically given by $|V_{ub}/V_{cb}|$.
We have to wait more accurate data to obtain more restrictive constraints.

At present the errors on the asymmetries are too large and do not really
provide stringent constraints. However, we include them here hoping that
they will be measured soon. By then one can easily include them in the fit to
obtain more stringent constraint on $\gamma$.

In Figure 5, we show the regions allowed by $\chi^2-\chi^2_{min}=1$
in the $\rho -\eta$ plane by the solid curve.
As mentioned that at present the constraint is
weak which can also be seen from Figure 6 where the minimal $\chi^2$
as a function
of $\gamma$ is shown by the curve (a) for the case with exact SU(3) symmetry,
although the $\chi^2_{min}$ per
degree of freedom is smaller than 1.
The 68\% allowed region is actually the same as that
from $|V_{ub}/V_{cb}|$
alone. However, when more precision data for rare charmless $B\to PP$ become
available, the restriction will become more stringent. For example, if the
error bars for all the quantities are reduced by a factor of 2.45, then the
regions in Figure 5 correspond to 95\% C.L. allowed regions.

SU(3) may not be an exact symmetry for $B\to PP$.
We now estimate SU(3) breaking effects.
The amplitudes $C_i$ for $B\to \pi\pi$ and $B\to K \pi$ will be
different if SU(3) is broken.
At present it is not possible to calculate the  breaking effects. To
have some idea about the size of the SU(3) breaking effects, we
work with the factorization estimate.
To leading order
the relation between the amplitudes for $B\to \pi\pi$ decays
$C_i(\pi\pi)$ and the amplitudes for $B\to K \pi$ decays
$C_i(K \pi)$ can be parameterized as $C_i(K \pi) = r C_i(\pi\pi)$,
and $r$ is approximately given by

\begin{eqnarray}
r \approx  {f_K\over f_\pi} = 1.22.
\end{eqnarray}
Here we have assumed that
the SU(3) breaking effects in $f_i$ and $F_0^{B\to i}$
are similar in magnitudes,
that is, $f_K/f_\pi \approx F^{B\to K}_0/ F^{B\to \pi}_0$\cite{30}.
Using the above to represent SU(3) breaking effect, we can obtain
another set of fitting results. They are shown in Figures 5 (dashed curve)
and 6 (curve (b)).  The best fit values for the amplitudes are

\begin{eqnarray}
&&C^P_{\bar 3} = 0.11,\;\;C^T_{\bar 3} = 0.33,\;\;
C^T_6= 0.22,\;\;C^T_{\overline {15}} = 0.18,\nonumber\\
&&\delta_{\bar 3} = 57^\circ,\;\; \delta_{6} = 200^\circ,\;\;
\delta_{\overline {15}} = 85^\circ.
\end{eqnarray}

The best fit values for $\rho$, $\eta$ and $\gamma$ are given by

\begin{eqnarray}
\rho = -0.39,\;\;
\eta = 0.07,\;\;
\gamma = 170^\circ.
\end{eqnarray}

In both exact and broken SU(3) cases, there are two local minimal in the
$\chi^2$ vs. $\gamma$ diagrams. The corresponding values of $\gamma$ are
very different with one of them around $87^\circ$ and another $170^\circ$.
These best fit values are dramatically different that those obtained
in section II. However the best fit values here can not be taken too seriously
because, as can be seen from Fig. 5, that at 39\% C.L. level, almost all
allowed range by $|V_{ub}/V_{cb}|$ is allowed by $B\to PP$ data.
At 68\% C.L. level, all allowed region by $|V_{ub}/V_{cb}|$ is allowed by 
data from $B\to PP$ decays.
Inconsistence
between $\gamma$ obtained in Section II and this section can not be
established. We have to wait more precise data on $B\to PP$ to decide.

In the literature it has often been quoted that $O_{1,2}$ do not contribute
to $B^- \to \bar K^0 \pi^-$ and therefore
$Br(B^- \to \bar K^0 \pi^-) = Br(B^+ \to
K^0 \pi^+)$. In the SU(3) language used here, this implies
$C = C^T_{\bar 3}e^{i\delta_{\bar 3}}
- C^T_6 e^{i\delta_6} - C^T_{\overline{15}} e^{i\delta_{\overline{15}}}
=0$.
This result has been used to derive several methods to determine
the phase $\gamma$.
We stress that this is not a result from SU(3) consideration and need to
be checked. For this reason
we also carried out analyses with the condition
$C=0$.
For this case, the minimal $\chi^2$ as a function of
$\gamma$ are also shown in Figure 6 (curves (a1) and (b1)) with exact SU(3)
and with
SU(3) breaking effects, respectively.
The best fit values with exact SU(3) symmetry are

\begin{eqnarray}
&&\rho = 0.05,\;\;\eta = 0.41,\;\;\gamma = 83^\circ,\nonumber\\
&&C^P_{\bar 3} = 0.13,\;\; C^T_{\bar 3} = 0.26,\;\;
C^T_6 = 0.17,\;\; C^T_{\overline{15}} = 0.16,\nonumber\\
&&\delta_{\bar 3} = -13^\circ,\;\;\delta_6 = -49^\circ,
\;\;\delta_{15} = 27^\circ.
\end{eqnarray}

And the best fit values with SU(3) breaking effects are
\begin{eqnarray}
&&\rho = 0.12,\;\;\eta = 0.39,\;\;\gamma = 73^\circ,\nonumber\\
&&C^P_{\bar 3} = 0.11,\;\; C^T_{\bar 3} = 0.24,\;\;
C^T_6 = 0.15,\;\; C^T_{\overline{15}} = 0.16,\nonumber\\
&&\delta_{\bar 3} = -15^\circ,\;\;\delta_6 = -56^\circ,\;\;
\delta_{\overline{15}} = 23^\circ.
\end{eqnarray}

The imposition of $C=0$
does not force these coefficients to be real. In order to get $C$
to be zero, the real and imaginary parts both have to cancel to satisfy the
condition.
The implications of this analysis will be discussed later.

\section{Combined fit}

In this section we carry out a combined fit of the sections II and III.
The total $\chi^2(total)$ is the sum of the $\chi^2(\mbox{II})$
with $\sin2\beta$ data included from section II plus
the $\chi^2(\mbox{III})$. Here $\chi^2(\mbox{III})$ is the $\chi^2$ in Eq. (36) of 
section III with $\chi^2(A, |V_{ub}/V_{cb}|)$ subtracted. 
This is because that $\chi^2(\mbox{II})$ already included
information from $A$ and $|V_{ub}/V_{cb}|$. The
results are shown in Figures 7 and 8. Since $\chi^2(\mbox{II})$ has a sharper 
dependence on $\gamma$ compared with $\chi^2(\mbox{III})$, 
the best fit values and
errors are dominantly determined by constraints in section II.

\begin{figure}[htb]
\centerline{
\DESepsf(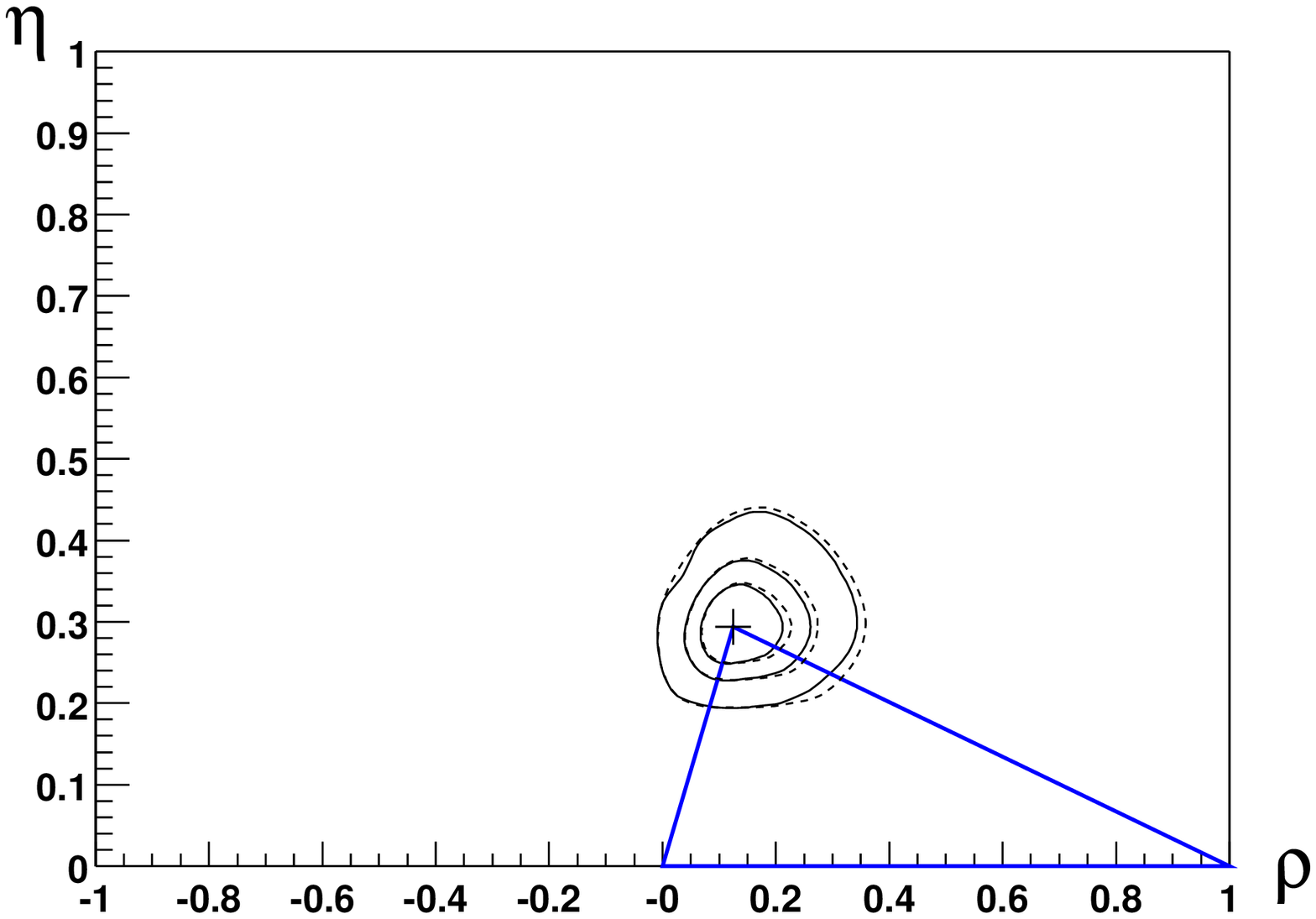 width 8cm)
\DESepsf(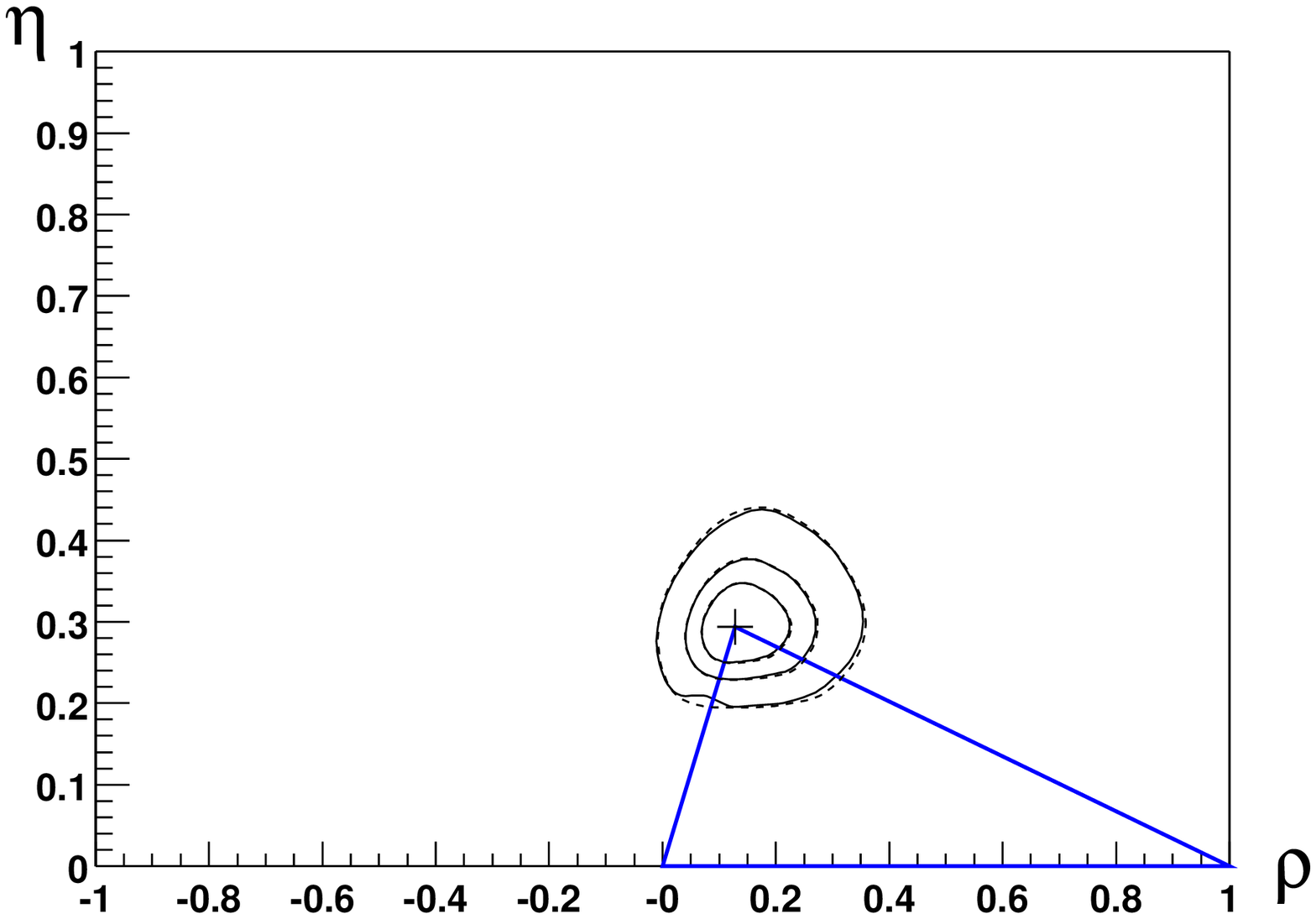 width 8cm)
}
\smallskip
\caption {Constraints on $\rho$ and $\eta$ using combined data
from $|\epsilon_K|$, $\Delta m_{B_{d,s}}$, $|V_{ub}/V_{cb}|$, $\sin2\beta$ and
$B\to \pi\pi$ and $B\to K\pi$.
The three regions from smaller to larger corresponds to
$\chi^2 - \chi^2_{min} =1$
allowed region which is at the 39\% C.L.,
the 68\% C.L. allowed region and
the 95\% C.L. allowed region, respectively. The figure on the left
is for the case with exact SU(3) and the one on the right is for the
case with SU(3) breaking effects. The dotted curves are for
fit in section II and the solid curves are for the combined fit.
}\label{fit-comb}
\end{figure}

\begin{figure}[htb]
\centerline{ \DESepsf(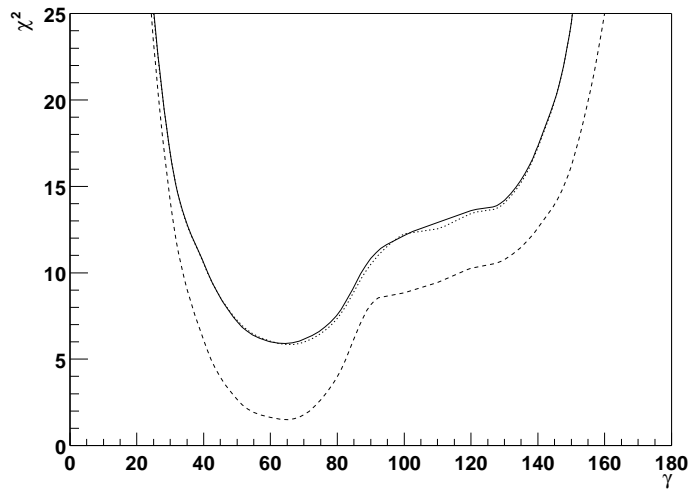 width 10cm)}
\smallskip
\caption {
$\chi^2$ as a function of $\gamma$
using combined data from $|\epsilon_K|$,
$\Delta m_{B_{d,s}}$, $|V_{ub}/V_{cb}|$, $\sin2\beta$ and rare $B\to PP$ data.
The dotted and solid
curves are for the fit with exact SU(3) and with SU(3) breaking effects, 
respectively. The dashed curve is the same as that from section II with
$\sin2\beta$ included.
}
\end{figure}

The best combined fit values with exact SU(3) symmetry for the hadronic
parameters are

\begin{eqnarray}
&&C^P_{\bar 3} = 0.13,\;\; C^T_{\bar 3} = 0.29,\;\;
C^T_6 = 0.16,\;\; C^T_{\overline {15}} = 0.20,\nonumber\\
&&\delta_{\bar 3} = -42^\circ,\;\;\delta_6 = -20^\circ,\;\;\delta_{\overline
{15}} = 35^\circ.
\end{eqnarray}
In the above we have not given errors for the hadronic parameters because
the constraints on them are weak.

The best fit values for $\rho$, $\eta$ and $\gamma$ and their 68\% C.L. errors
are given by

\begin{eqnarray}
\rho = 0.12^{+0.09}_{-0.05},\;\;\eta = 0.29^{+0.06}_{-0.04},
\;\;\gamma = 67^{+10^\circ}_{-13^\circ},
\end{eqnarray}
and the 95\% C.L. allowed ranges for $\rho$, $\eta$ and $\gamma$ are

\begin{eqnarray}
&&0.01 < \rho < 0.29,\;\; 0.21< \eta < 0.40,\nonumber\\
&&-0.86 < \sin 2\alpha < 0.45, \;\;0.45 < \sin 2\beta < 0.79,\;\;
43^\circ < \gamma < 87^\circ.
\end{eqnarray}

The best combined fit values with SU(3) breaking effects for the hadronic
parameters are
\begin{eqnarray}
&&C^P_{\bar 3} = 0.11,\;\; C^T_{\bar 3} = 0.29,\;\;
C^T_6 = 0.16,\;\; C^T_{\overline {15}} = 0.20,\nonumber\\
&&\delta_{\bar 3} = -33^\circ,\;\;\delta_6 = -40^\circ,\;\;\delta_{
\overline {15}} = 31^\circ.
\end{eqnarray}

\begin{eqnarray}
&&\rho = 0.13^{+0.09}_{-0.06},\;\;\eta = 0.29^{+0.06}_{-0.04},
\;\;\gamma = 66^{+10^\circ}_{-13^\circ},
\end{eqnarray}
and the 95\% C.L. allowed ranges for $\rho$, $\eta$ and $\gamma$ are

\begin{eqnarray}
&&0.01 < \rho < 0.30,\;\; 0.21< \eta < 0.41,\nonumber\\
&&-0.87 < \sin 2\alpha < 0.44, \;\;0.46 < \sin 2\beta < 0.80,\;\;
42^\circ < \gamma < 87^\circ.
\end{eqnarray}

\section{Discussions and Conclusions}

At present the rare charmless hadronic $B$
decay data have large error bars.
The main contribution to the $\chi^2$ for the analysis in sections III and IV
come from the branching ratio for $\bar B^0 \to \bar K^0 \pi^0$. In the
cases discussed, this mode alone contribute about 2.5 to the $\chi^2$. The
best fit value of the branching ratio
is only about half of the experimental central value.
We suspect that there may be some systematic errors in the measurement of
this branching ratio.
If the present central value persists, it may be an indication of badly broken
SU(3) symmetry or new physics beyond the SM. It is important to improve the
precision of experimental data to decide whether new physics is needed.

Because of the large error bars associated with the $B \to PP$ data,
the ranges determined for
the related parameters have large error bars. Especially the phase $\gamma$
has a large range allowed using the $B\to PP$ data alone. However, the fit
shows no conflict between the fit from the consideration using
$|\epsilon_K|$, $\Delta m_{B_{d,s}}$, $|V_{ub}/V_{cb}|$ and $\sin2\beta$ data.
Future experimental data will be able to provide a more accurate determination
of the phase $\gamma$.

Before closing we
would like to make a few comments about our analysis and some other related
calculations. Our first comment concerns the general SU(3) analysis and
factorization calculations.

Assuming factorization and SU(3) symmetry, that is the decay constants for
all octet pseudoscalars $P$ are equal, and the form
factors for $B\to P$ are also equal, one obtains\cite{24}

\begin{eqnarray}
 &&C_{\bar 3}^T=\{\frac{3 a_2-a_1}{8}-[(a^{uc}_4+a_6^{uc} R)
+\frac{3}{16}(a_7^{uc}-a_9^{uc})+\frac{1}{16}(a_{10}^{uc}+a_8^{uc} R)]\}
 X \nonumber\\
 &&C_6^T=\{\frac{a_2-a_1}{4}-\frac{3}{8}(a_{10}^{uc}-a_9^{uc}+a_7^{uc}+a_8^{uc} R)\}
 X \nonumber\\
 &&C_{\overline {15}}^T=\{\frac{a_1+a_2}{8}-\frac{3}{16}(a_9^{uc}+a_{10}^{uc}-a_7^{uc}
+a_8^{uc} R)\}
 X
\label{facteq}
\end{eqnarray}
where $X=f_\pi F^{B \to \pi}_0(m_\pi^2)(m_B^2-m_\pi^2)$ and
$R=m^2_{K}/(m_b-m_q)(m_s+m_q)$.
These amplitudes are related to the  ``tree'' contributions, $a^{uc}_i
=a_i^u-a_i^c$ with $a_{2i} = c_{2i}+c_{2i-1}/N$ and $a_{2i-1} = c_{2i-1}
+c_{2i}/N$.

The penguin amplitudes are given by\cite{24}

\begin{eqnarray}
 &&C_{\bar 3}^P=-[(a_4^{tc}+a_6^{tc} R)+\frac{3}{16}(a_7^{tc}-a_9^{tc})
+\frac{1}{16}(a_{10}^{tc}+a_8^{tc} R)]X\;, \nonumber\\
 &&C_6^P=-\frac{3}{8}(a_{10}^{tc}-a_9^{tc}+a_7^{tc}+a_8^{tc} R)X\;, \nonumber\\
 &&C_{\overline {15}}^P=-\frac{3}{16}(a_9^{tc}+a_{10}^{tc}-a_7^{tc}+a_8^{tc} R)X\;.
\end{eqnarray}
where $a^{tc}_i =a_i^t-a_i^c$.
When small contributions from $a_{7,8}^{ij}$
terms are neglected, one recovers the relations in Eq. (\ref{relation}).
Numerically, we have

\begin{eqnarray}
&&C_{\bar 3}^P=0.09,\;\;
C_{\bar 3}^T=0.42, \;\;
C_6^T=0.26, \;\;
C_{\overline{15}}^T=0.15, \nonumber\\
&&\delta_{\bar 3} = -15.7^\circ,\;\;
\delta_{6} = -14.5^\circ,\;\;
\delta_{\overline{15}} = -14.5^\circ.
\end{eqnarray}
In the above we have used the convention with $C^P_{\bar 3}$ to be real.

The amplitudes are in the same order of magnitude as the best fit values in
sections III and IV, but the phase can be
very different. In the factorization approximation calculation here, phases
are only due to short distance interaction, rescattering of quarks. Long
distance contributions can change these phases. The results of the best fit
values for the phases indicating that there may be
large long distance rescattering
effects.

Our second comments concerns the combination of the
SU(3) invariant decay amplitude
$C = C_{\bar 3}^Te^{i\delta_{\bar 3}} - C_6^Te^{i\delta_6} -
C_{\overline{15}}^Te^{\delta_{\overline{15}}}$.
It has been usually assumed that in the literature that
$C = 0$. This leads to $Br(B^+\to K^0 \pi^+) =
Br(B^- \to \bar K^0 \pi^-)$. This result played a crucial role in
several methods to
constrain and to determine the phase $\gamma$, for example,
using\cite{8} $B^-\to K^- \pi^0,
\bar K^0 \pi^-, \pi^- \pi^-$, $\bar B^0 (B^-) \to K^+ \pi^-, \pi^+\pi^-
(\bar K^0 \pi^-)$, and $B^- \to K^- \pi^0, \bar K^0 \pi^-, K^- \eta$.

We point out that $C=0$
is based on factorization calculation neglecting annihilation
contributions and also penguin contributions\cite{24}.
In fact, using factorization calculation when penguin contributions are included,
$C$ does not equal to zero, but
$C = C^T_{\bar 3}(penguin)$.
$C^T_{\bar 3}(penguin)$ can be obtained from Eq.(\ref{facteq}) (the terms
proportional to $c^{uc}_i$ in $C^T_{\bar 3}$.
In the factorization framework, we can easily check whether $C=0$ is a
good approximation. Using the result in
Eq. (\ref{facteq}) we find that
the $|C^T_{\bar 3}(penguin)/C^T_{\bar 3}|$ is of order 5\%.
It is therefore reasonable to assume
the penguin contribution to be small and $C\approx 0$.

One should also be aware that when going beyond factorization approximation
and include rescattering effects $C$ may deviate from zero. It
should be tested. The fitting program proposed in this paper can be easily used
to achieve this goal. From the best fit values in the previous sections,
we clearly see that $C$ can easily deviate from zero. For example, in the case
with exact SU(3), the best fit value using rare $B$ decay data $C$ is,
$C = 0.05-i0.20$ and with SU(3) breaking effects $C$ is, $C =0.37+ i0.18$ which
are the same order of magnitude as individual $C^T_i$.
One needs more data to
achieve a better test. Until then, the use of the methods based on the above
equation have to be treated with caution.

Our final comments concerns the uncertainties in the present analysis.
In this paper we have developed a method based on SU(3) flavor symmetry to
determine the CP violating phase $\gamma$. We find that when annihilation
contributions are neglected, there are only seven hadronic parameters in the
SM related to $B\to PP$ decays. The annihilation contributions
are small is based on factorization approximation. If it turns out that they
are not small, as some model calculations indicated that the penguin related
annihilation contribution $A_{\bar 3}^P$ may be sizeable,
one needs to include it into the analysis.
However, from Table II one can see that $A^P_{\bar 3}$ does not show up in
$B\to K \pi$ decays, but only to $B\to \pi\pi$ which is suppressed by small
Wilson coefficients.
One can also carried out an analysis including $A^P_{\bar 3}$ into the fit
when more experimental data become available. Future experimental data with
better accuracy will provide more information.

In the estimate of SU(3) breaking effects, we have parameterized the
SU(3) breaking effects in a simple form with $C_i(K\pi) = (f_K/f_\pi)
C_i(\pi \pi)$. In general the SU(3) breaking effects may be more complicated.
More systematic study of SU(3) breaking effects are needed in order to obtain
more accurate determination of the phase $\gamma$.
But in any rate we hope that the method developed here will
help to provide useful information about the hadronic matrix elements
and also the CP violating phase $\gamma$.

In conclusion, in this paper we have developed a method to determine the CP
violating phase $\gamma$ based on the flavor SU(3) symmetry.
We find that present data
can already give some constraint on $\gamma$ and it is consistent
with the
constraint obtained by using $|\epsilon_K|$, $\Delta m_{B_{d,s}}$,
$|V_{ub}/V_{cb}|$ and $\sin2\beta$ data.
We also carried out an analysis combining
data from $\epsilon_K$, $\Delta m_{B_{d,s}}$,
$|V_{ub}/V_{cb}|$, $\sin2\beta$ 
and data from rare charmless hadronic $B$ decays. The
combined analysis gives
$\gamma=67^\circ$ for the best fit value and $43^\circ \sim 87^\circ$
as the 95\% C.L. allowed range. Although there are uncertainties in the
fit program, the method developed in the present paper
can provide useful information about the hadronic matrix elements
for rare charmless hadronic B decays and the CP violating phase $\gamma$.

XGH would like to thank P. Chang and D. London for useful discussions.
XGH, YKH and JQS were
supported by the National Science Council of ROC under grant
number NSC89-2112-M-002-058 and NCTS of ROC,
and, YLW and YFZ were supported by the
NSF of China under the grant No. 19625514.


\begin{references}

\bibitem{1}  M. Kobayashi
and T. Maskawa, Prog. Theor. Phys. {\bf 49}, 652(1973).

\bibitem{2}  N. Cabibbo, Phys. Rev. Lett. {\bf 10}, 531(1963).

\bibitem{3} Particle Data Group, Eur. Phys. J. {\bf C 15} 1(2000).

\bibitem{4} L. Wolfenstein, Phys. Rev. Lett. {\bf 51}, 1945(1983).

\bibitem{5}  S. Mele, Phys.Rev. {\bf D59} 113011(1999);
A. Ali, D. London, Eur.Phys.J. {\bf C9} 687(1999).

\bibitem{5a}
F. Parodi, P. Roudeau and A. Stocchi, Nuovo. Cim. {\bf A112}, 833(1999);
F. Caravaglios et al., e-print hep-ph/0002171.

\bibitem{6} CLEO Collaboration, D. Cronin-Hennessy et al., Phys. Rev.
Lett.{\bf 85}, 525(2000).

\bibitem{7}
N. G. Deshpande, et al.,
Phys. Rev. Lett. {\bf 82} 2240(1999);
X.-G. He, W.-S. Hou and K.-C. Yang, Phys. Rev. Lett.
{\bf 83}, 1100(1999);
W.-S. Hou, J. G. Smith, F. Wurthwein, e-print hep-ex/9910014;
C. Isola and T.N. Pham, e-print hep-ph/0009210;
Y-L. Wu and Y.-F. Zhou, Phys. Rev. {\bf D62}, 036007(2000);
A. Buras and R. Fleischer, Eur. Phys. J. {\bf C16}, 97(2000);
M. Beneke et al., e-print hep-ph/0007256; Y.-Y. Keum, H.-n. Li and
A.I. Sanda, e-print hep-ph/0004004.

\bibitem{8} R. Fleischer and T. Mannel, Phys. Rev. {\bf D57}, 2752(1998);
M. Neubert and J. Rosner, Phys. Lett. {\bf B441}, 403(1998);
M. Neubert and J. Rosner, Phys. Rev. Lett. {\bf 81}, 5076(1998);
X.-G. He, C.-L. Hsueh and J.-Q. Shi, Phys. Rev. Lett.
{\bf 84} 18(2000); M. Gronau and J. Rosner, Phys. Rev. {\bf D57}, 6843(1998);
N.G. Deshpande and X.-G. He, Phys. Rev. Lett. {\bf 75}, 3064(1995).

\bibitem{9}
M. Savage and M. Wise, Phys. Rev.
{\bf D39}, 3346(1989); $ibid$ {\bf D40}, Erratum, 3127(1989);
X.-G. He, Eur. Phys. J. {\bf C9}, 443(1999);
N. G. Deshpande, X.-G. He, and J.-Q. Shi, Phys. Rev. {\bf D62}, 034018(2000).


\bibitem{10} M. Gronau et al., Phys. Rev. {\bf D50}, 4529 (1994); {\bf D52},
6356 (1995); $ibid$, 6374 (1995);
A.S. Dighe, M. Gronau and J. Rosner, Phys. Rev. Lett. {\bf 79}, 4333 (1997); L.L. Chau et al.,
Phys. Rev. {\bf D43}, 2176 (1991); D. Zeppendfeld, Z. Phys. {\bf C8}, 77(1981).

\bibitem{11} M. Gronau and D. London, Phys. Rev. Lett. {\bf 65},
3381(1990); N. G. Deshpande and X.-G. He,
Phys. Rev. Lett. {\bf 75}, 1703(1995); M. Gronau and J. Rosner,
Phys. Rev. {\bf D57}, 6843(1998); $ibid$, {\bf D61}, 073008(2000);
Phys. Lett. {\bf B482}, 71(2000), M. Gronau, D. Pirjol and T.-M. Yan,
Phys. Rev. {\bf D60}, 034021(1999).

\bibitem{12} X.-G. He, J.-Y. Leou and C.-Y. Wu, Phys. Rev. {\bf D62},
114015(2000).

\bibitem{13} Y.-F. Zhou et al., e-print hep-ph/0006225
(in press in Phys. Rev. D).

\bibitem{14} S. Herrlich and U. Nierste, Nucl. Phys.{\bf B 419}, 292(1994);
A. J. Buras, M. Jamin and P.H. Weisz, Nucl. Phys. {\bf B 347}, 491(1990);
S. Herrlich and U. Nierste, Phys. Rev. {\bf D 52}, 6505(1995).

\bibitem{15} T. Draper, e-print hep-lat/9810065, Nucl. Phys. Proc. Suppl.
{\bf 73}, 43(1999); S. Sharpe, e-print
hep-lat/9811006;
JLQCD Collab.,
S. Aoki et al., Nucl. Phys. B (proc. Suppl) {\bf 63A-C} 281(1998).


\bibitem{16} A. Ali and D. London, Z. Phys. {\bf C 65} 431(1995);
Takeo Inami, C.S. Lim, B. Takeuchi, M. Tanabashi, Phys. Lett. {\bf B381} 458(1996).

\bibitem{17}
The LEP B oscillation WG, http://lepbosc.web.cern.ch/LEPBOSC, updated for
XXXth International Conference on High Energy Physics, Osaka,
Japan.

\bibitem{18} J. M. Flynn, C. T. Sachrajda, hep-lat/9710057,
to appear in Heavy Flavours (2nd edition)
edited by A J Buras and M Lindner (World Scientific, Singapore).

\bibitem{bsmixing} H. G. Moser and A. Roussani, Nucl. Inst. Meth. {\bf A384},
491(1997).

\bibitem{19} 
The BABAR Collaboration, B. Aubert, et al,  submitted to P.R.L.
A. Abashian et al. (Belle Collaboration), submitted to P.R.L.



\bibitem{20} T. Affolder et al. (CDF Collaboration), Phys. Rev. {\bf D61}
072205(2000).

\bibitem{21} The ALEPH Collaboration, e-print hep-ex/0009058, Phys. Lett.
{\bf 492}, 259(2000).

\bibitem{22} G. Buchalla, A. Buras and
M.Lautenbacher, Rev. Mod. Phys. {\bf 68}, 1125(1996);
M.Ciuchini et.al., Nucl. Phys. {\bf B415}, 403(1994).

\bibitem{23} N. G. Deshpande and X.-G. He, Phys. Lett. {\bf B336}, 471 (1994).


\bibitem{24} A. Ali, G. Kramer and C.-D. Lu, Phys. Rev. {\bf D58}, 094009(1998);
$ibid$ {\bf D59} 014005(1999);
A. Datta, X.-G. He and S. Pakvasa,
Phys. lett. {\bf B419}, 369(1998);
Y.-H. Chen, H.-Y. Cheng, B. Tseng, K.-C. Yang,
Phys. Rev. {\bf D60} 094014(1999).

\bibitem{25}
 R. Fleischer, Z. Phys. {\bf C62}, 81(1994);
Phys. Lett. {\bf B321}, 259(1994); Phys. Lett. {\bf B332}, 419(1994);
N. Deshpande, X.-G. He and J. Trampetic,
Phys. Lett. {\bf B345}, 547 (1995); N.G. Deshpande and X.-G. He,
Phys. Rev. Lett. {\bf 74}, 26 (1995); 4099(E) (1995).

\bibitem{26} D. Urner, CLEO TALK 00-33, DPF2000, Columbus, Ohio, USA,
August 9-12, 2000;
R. Stroynowshi, CLEO Talk 00-30, ICHEP2000, Osaka, Japan, July 27 to August
2, 2000.

\bibitem{27} 
T. Iijima, Belle Collaboration, Talk at BCP4, Ise-Shima, Japan, 
February 19-23, 2001.

\bibitem{28} 
A. Hoecker, Paris-Sud ,  Babar Collaboration, Talk at BCP4, Ise-Shima, 
Japan,
February 19-23, 2001.

\bibitem{29} CLEO Collaboration, S. Chen et al., Phys. Rev. Lett.{\bf85},
525(2000).

\bibitem{30} D. Melikhov and B. Stech, Phys. Rev. {\bf D62}, 014006(2000).

\end{references}
\end{document}